\documentclass[usenatbib,12pt, preprint]{aastex6}
\usepackage{natbib}

\slugcomment{Version: \today}

\usepackage{color}

\begin{document}
\title{The JCMT Transient Survey: Stochastic and Secular Variability of Protostars and Disks In the Sub-Millimeter Observed Over Eighteen Months}
%\shorttitle{}
  \author{
Doug Johnstone\altaffilmark{1,2}, 
Gregory J.\ Herczeg\altaffilmark{3}, 
Steve Mairs\altaffilmark{1,2,4}, 
Jennifer Hatchell\altaffilmark{5}, 
Geoffrey C.\ Bower\altaffilmark{6}, 
Helen Kirk\altaffilmark{1},
%
%DR subteam:
%
James Lane\altaffilmark{2},
Graham S.\ Bell\altaffilmark{4},
Sarah Graves\altaffilmark{4},
%
%Regions Sub-team/Co-I’s
%
Yuri Aikawa\altaffilmark{7}
Huei-Ru Vivien Chen\altaffilmark{8},
Wen-Ping Chen\altaffilmark{9},
Miju Kang\altaffilmark{10},
Sung-Ju Kang\altaffilmark{10},
Jeong-Eun Lee\altaffilmark{11},
Oscar Morata\altaffilmark{12},
Andy Pon\altaffilmark{13},
Peter Scicluna\altaffilmark{14},
Aleks Scholz\altaffilmark{15},
Satoko Takahashi\altaffilmark{16,17},
Hyunju Yoo\altaffilmark{11,18},
and The JCMT Transient Team
}

\altaffiltext{1}{NRC Herzberg Astronomy and Astrophysics, 5071 West Saanich Rd, Victoria, BC, V9E 2E7, Canada}
\altaffiltext{2}{Department of Physics and Astronomy, University of Victoria, Victoria, BC, V8P 5C2, Canada}
\altaffiltext{3}{Kavli Institute for Astronomy and Astrophysics, Peking University, Yiheyuan 5, Haidian Qu, 100871 Beijing, People's Republic of China}
\altaffiltext{4}{East Asian Observatory, 660 North A`oh\={o}k\={u} Place, University Park, Hilo, Hawaii 96720, USA}
\altaffiltext{5}{Pysics and Astronomy, Exeter University, Stocker Road, Exeter EX4 4QL, UK}
\altaffiltext{6}{Academia Sinica Institute of Astronomy and Astrophysics, 645 N.\ A'ohoku Place, Hilo, HI 96720, USA}
\altaffiltext{7}{Department of Astronomy, University of Tokyo, Tokyo, Japan}
\altaffiltext{8}{Department of Physics and Institute of Astronomy, National Tsing Hua University, Taiwan}
\altaffiltext{9}{Graduate Institute of Astronomy, National Central University, 300 Jhongda Road, Zhongli, Taoyuan, Taiwan}
\altaffiltext{10}{Korea Astronomy and Space Science Institute, 776 Daedeokdae-ro, Yuseong-gu, Daejeon 34055, Republic of Korea}
\altaffiltext{11}{School of Space Research, Kyung Hee University, 1732, Deogyeong-Daero, Giheung-gu Yongin-shi, Gyunggi-do 17104, Korea}
\altaffiltext{12}{Academia Sinica Institute of Astronomy and Astrophysics, 11F of AS/NTU Astronomy-Mathematics Building, No.1, Sec. 4, Roosevelt Rd, Taipei 10617, Taiwan}
\altaffiltext{13}{Department of Physics and Astronomy, The University of Western Ontario, 1151 Richmond Street, London, ON, N6A 3K7, Canada}
\altaffiltext{14}{Academia Sinica Institute of Astronomy and Astrophysics, P. O. Box 23-141, Taipei 10617, Taiwan}
\altaffiltext{15}{SUPA, School of Physics \& Astronomy, North Haugh, St Andrews, KY16 9SS, United Kingdom}
\altaffiltext{16}{Joint ALMA Observatory, Alonso de C\'ordova 3107, Vitacura, Santiago, Chile}
\altaffiltext{17}{National Astronomical Observatory of Japa, 2-21-1 Osawa, Mitaka, Tokyo 181-8588, Japan}
\altaffiltext{18}{Department of Astronomy and Space Science, Chungnam National University, 99 Daehak-ro, Yuseong-gu, Daejeon 34134, Republic of Korea}

\begin{abstract}
We analyze results from the first eighteen months of monthly sub-mm monitoring of eight star-forming regions in the JCMT Transient Survey.  In our search for stochastic variability in 1643 bright peaks, only the previously identified source, EC\,53, shows behavior well above the expected measurement uncertainty.  Another four sources, two disks and two protostars, show moderately-enhanced standard deviations in brightness, as expected for stochastic variables.
For the two protostars, this apparent variability is the result of single epochs that are much brighter than the mean.   
In our search for secular brightness variations that are linear in time, we measure the fractional brightness change per year for 150 bright peaks, fifty of which are protostellar.  The ensemble distribution of slopes is well fit by a normal distribution with $\sigma \sim 0.023$.  Most sources are not rapidly brightening or fading in the sub-mm. Comparison against time-randomized realizations shows that the width of the distribution is dominated by the uncertainty in the individual brightness measurements of the sources. A toy model for secular variability reveals that an underlying Gaussian distribution of linear fractional brightness change $\sigma = 0.005$ would be unobservable in the present sample, whereas an underlying distribution with $\sigma = 0.02$ is ruled out. Five protostellar sources, 10\% of the protostellar sample, are found to have robust secular measures deviating from a constant flux. The sensitivity to secular brightness variations will improve significantly with a larger time sample, with a factor of two improvement expected by the conclusion of our 36-month survey.
\end{abstract}


\section{Introduction}
\label{Sec:Intro}

A protostellar core begins its life by growing smoothly from a collapsing envelope.  Protoplanetary disks are thought to form early in the protostellar lifetime \citep[e.g.][]{jorgensen08} and, once formed, the disk channels accretion from the envelope onto the star 
\citep[e.g.][]{hartmann97}.  While the initial protostellar growth from the envelope should be steady, disk accretion is expected to be variable because instabilities are expected to quickly grow \citep[see review by][]{armitage15}.  

These instabilities produce macroscopic structures that are detectable even in young disks, such as rings in HL Tau \citep{hltau}, spiral density waves in Elias 2-27 \citep{perez16}, and multiplicity from disk fragmentation in L1448 IRS3B \citep{tobin16triple}.  However, while such structures are detectable at large radii, the identification of structures in inner disks is typically beyond the detection limits of current instrumentation.  The size and scale of instabilities in the inner disk may instead be indirectly traced by monitoring accretion from the disk onto the star.  Variability in accretion is commonly detected in optically-visible systems and has been used to infer the 
presence of instabilities in the disk \citep[see reviews by][]{audard14,hartmann16}.
Small, short-lived flickers seen in high-cadence monitoring \citep[e.g.][]{cody17} and spectroscopic campaigns \citep[e.g.][]{costigan14} suggests a non-steady star-disk connection \citep[e.g.][]{romanova12}.  Months-long bursts of EXor systems may reveal the expansion and contraction of the magnetospheric cavity \citep{dangelo10}.  The largest known bursts, seen as FUor objects, are caused by a factor of $10^4$ increase in the accretion rate and may last for over a century \citep[e.g.][]{zhu09}. {Recently, \citet{liu2017} surveyed a sample of 29 FUors and Exors at 1.3\, mm with the SMA and tentatively detected two sources, V2494\,Cyg and V2495\,Cyg, with 30-60\% millimeter flux variability on year timescales.}

This accretion variability, and in particular large bursts, is thought to play an important role in the chemical evolution of the envelope and disk \citep[e.g.][]{kim12,harsono15,frimann17} as well as the contraction rate of the star \citep[e.g.][]{hosokawa11,baraffe17}.  However, at younger stages of evolution, the star/disk system is obscured by an optically-thick envelope that enshrouds the accretion, preventing direct detection of accretion and therefore any accretion variability.   Yet during this period, the star accretes most of its mass, and the effects of any variability are expected to be the most significant.

\citet{johnstone13} examined the variability of sub-mm dust continuum emission as a new method to probe accretion variability onto the star. They noted that the variable mass accretion $\dot M_a(t)$ with time, $t$, onto deeply embedded protostars should be observable through the proxy measurement of accretion luminosity, $L_a(t) \propto \dot M_a(t)$, and that this varying accretion luminosity should leave a signature on the protostellar envelope. Since the heat capacity of dust is small, the absorption and re-emission from dust within the dense envelope should quickly bring the spectral energy distribution (SED) of the protostellar core into an equilibrium determined by the new accretion luminosity. Indeed, the only relevant time delay was found to be the light-crossing time, of order days to months depending on the size of the envelope and the wavelength of interest. The longest delay times are associated with single-dish sub-mm observations since, for these observations, the change in the temperature of the outer envelope is responsible for the majority of the observed change in emission. %Furthermore, \citet{johnstone13} noted that at sub-mm wavelengths the brightness change is often proportional to the temperature change \citep[see also][]{yoo17}. As such, the amplitude of the signal should roughly scale as $L_a(t)^{1/4}$. 
The change in the temperature of the gas proceeds much slower than for the dust, because the gas has a much higher heat capacity and is warmed (or cooled) primarily through collisions with the dust.

To search for variability in sub-mm dust emission, "The JCMT Transient Survey" \citep{herczeg17} is monitoring eight star-forming regions with the SCUBA-2 \citep{holland13} sub-mm bolometer on the James Clerk Maxwell Telescope (JCMT).  These star-forming regions are imaged with an approximately monthly cadence in order to search for indicators of variability. Our survey was begun in December 2015 and will run through at least January 2019. Each region has a sufficient number of compact bright sub-mm sources to allow for significant improvement in the nominal pointing and brightness calibration \citep{mairs17}. This survey has already uncovered the first robust detection of a sub-mm periodic variable \citep[EC 53,][]{yoo17} and, through comparison with previous Gould Belt Survey \citep[GBS;][]{ward-thompson07} observations of these same star-forming regions, has also revealed that a handful of sources have small, but robust variations in their brightness across two to four years \citep{mairs17b}.

In this paper we analyze the first eighteen months of the JCMT Transient Survey to search for evidence of stochastic and secular sub-mm variability within eight star-forming regions. In \S\ref{Sec:DR} we recap the data reduction and calibration procedures that are applied to the individual observations to ensure a reliable and uniform data product. Using deep, stacked images of each region we determine the location of robust sub-mm peaks and collate these sources with known protostars and disk sources. In \S\ref{Sec:Sigma} we analyze, for each identified source, the standard deviation in the brightness across all epochs and search for those sources which show significant variation from the expected value, a potential signpost of stochastic variable behavior. In \S\ref{Sec:Slope} we measure the degree of secular variability in the brightness observed for each source by fitting linear slopes in time to the brightness measurements. Consideration of the uncertainty in the slope measurement and comparison against time-randomized observations of the same sources allows for the determination of robust results. In \S\ref{Sec:Individual} we discuss individual sources in more detail, looking both at the robust and candidate sources uncovered in the previous two sections as well as potential variable sources obtained from the literature.  Section \ref{Sec:Discussion} places the observations obtained to date with the JCMT Transient Survey in context with the broader search for variability in both observations and numerical simulations. Finally, in \S\ref{Sec:Summary} we summarize our results.

\section{Data Reduction}
\label{Sec:DR}
The observations included in this paper were taken as part of the JCMT Transient Survey \citep{herczeg17}, with roughly monthly observations from 22 December 2015 through 16 June 2017. 
{The JCMT Transient Survey uses the Submillimetre Common User Bolometer Array 2 \citep[SCUBA-2;][]{holland13} to simultaneously image the sky at both \mbox{850 $\mu$m} and \mbox{450 $\mu$m} with effective beam FWHMs of 14.6\arcsec and 9.8\arcsec, respectively. In this paper, we focus only on the \mbox{850 $\mu$m} images. The \mbox{450 $\mu$m} data is far more susceptible to slight changes in the precipitable water vapour and this introduces complications to the data reduction and calibration procedures for a consistent comparison of peak brightnesses across many epochs. All of the observations were performed while the optical depth at 225 GHz ($\tau_{225}$) was less than 0.12 as measured by the JCMT water vapour Radiometer \citep{dempsey2008}. The integration time of the observations varied between 20 and 40 minutes, depending on weather conditions, in order to achieve a consistent sensitivity of \mbox{$\sim$10 mJy beam$^{-1}$} at \mbox{850 $\mu$m}.
} 
The \mbox{850 $\mu$m} data were reduced and calibrated following the procedure described by \citet{mairs17} using the iterative map-making software  {\sc{makemap}} (described in detail by \citealt{chapin2013}) in the {\sc{SMURF}} package (\citealt{jenness2013}) found within the {\sc{Starlink}} software \citep{currie2014}. Each \mbox{850 $\mu$m} map is gridded to 3$\arcsec$ pixels and convergence of the iterative solution is defined when the difference in individual pixels changes on average by $<$0.1\% of the rms noise present in the map. Emission on scales larger than $\sim 200\arcsec$ is filtered out of these maps while smaller-scale structures are robustly recovered \citep[see ][]{chapin2013,mairs15,mairs17}. While the \mbox{CO(J=3-2)} emission line contributes to the flux measured in \mbox{850 $\mu$m} continuum observations \citep{johnstone99,drabek2012,coude2016}, \cite{mairs2016} show that the peak brightnesses of compact sources are not significantly affected by the removal of this line {and therefore no attempt to remove the line has been undertaken for this analysis.}

The pertinent observation and calibration information is provided in Table\ \ref{Tab:Epochs}. The Transient Survey calibration procedure \citep{mairs17} improves on the default JCMT calibrations \citep{dempsey13} in two ways. First, image alignment is improved from the $2\arcsec - 6\arcsec$ telescope pointing uncertainty to a relative value per field of $< 1\arcsec$. Second, the default brightness calibration of $\sim 5\% - 10\%$ is lowered to a relative value per field per epoch of $\sim 2\%$ (for details refer to \citealt{mairs17}).

%chopped from body of text
\begin{deluxetable}{lllrrrrcr}
\tablecolumns{9}
\tablewidth{0pc}
\tablecaption{Regions, Epochs, and Calibration \label{Tab:Epochs}}
\tablehead{
\colhead{Region} &
\colhead{R.A.} &
\colhead{Dec.} &
\colhead{Epoch} & 
\colhead{Date}  & 
\colhead{Scan} &
\colhead{$\tau_{225}$\tablenotemark{a}} &
\colhead{Noise} &
\colhead{FCF\tablenotemark{b}}\\
\colhead{}&
\colhead{}&
\colhead{}&
\colhead{}&
\colhead{}&
\colhead{}&
\colhead{}&
\colhead{(Jy\,bm$^{-1}$)}&
\colhead{}
}
\startdata
IC\,348  & 3h 44m 18.00s & +32:04:59.00 & 1& 2015-12-22 & 00019 & 0.064 & 0.011 & 1.022\\
& & & 2& 2016-01-15 & 00022 & 0.072 & 0.009 & 1.050\\
& & & 3& 2016-02-05 & 00018 & 0.037 & 0.013 & 0.968\\
& & & 4& 2016--02-26 & 00020 & 0.054 & 0.012 & 0.973\\
& & & 5& 2016-03-18 & 00027 & 0.048 & 0.011 & 0.942\\
& & & 6& 2016-04-17 & 00009 & 0.036 & 0.011 & 0.927\\
& & & 7& 2016-08-26 & 00040 & 0.082 & 0.014 & 0.968\\
& & & 8& 2016-11-26 & 00022 & 0.049 & 0.010 & 1.226\\
& & & 9& 2017-02-09 & 00028 & 0.089 & 0.012 & 1.030\\
& & &10& 2017-03-20 & 00019 & 0.086 & 0.011 & 0.980\\
\hline
NGC\,1333& 3h 28m 54.00s & +31:16:52.00 & 1& 2015-12-22 & 00018 & 0.061 &  0.011 & 1.039\\
& & & 2& 2016-01-15 & 00010 & 0.081 & 0.012 & 0.978\\
& & & 3& 2016-02-05 & 00017 & 0.037 & 0.012 & 1.001\\
& & & 4& 2016-02-29 & 00017 & 0.042 & 0.011 & 1.001\\
& & & 5& 2016-03-25 & 00011 & 0.057 & 0.011 & 0.927\\
& & & 6& 2016-08-02 & 00031 & 0.094 & 0.012 & 1.027\\
& & & 7& 2016-08-30 & 00048 & 0.090 & 0.013 & 1.020\\  
& & & 8& 2016-11-19 & 00088 & 0.067 & 0.008 & 1.058\\
& & & 9& 2016-11-26 & 00021 & 0.048 & 0.010 & 1.164\\ 
& & &10& 2017-02-06 & 00029 & 0.124 & 0.013 & 0.948\\
& & &11& 2017-03-18 & 00015 & 0.111 & 0.014 & 0.948\\
\hline
NGC\,2024& 5h 41m 41.00s & -01:53:51.00 & 1& 2015-12-26 & 00049 & 0.117 & 0.013 & 0.975\\
& & & 2& 2016-01-16 & 00022 & 0.057 & 0.009 & 0.999\\
& & & 3& 2016-02-06 & 00013 & 0.043 & 0.011 & 1.046\\
& & & 4& 2016-02-29 & 00022 & 0.044 & 0.011 & 1.089\\
& & & 5& 2016-03-25 & 00021 & 0.057 & 0.012 & 0.935\\
& & & 6& 2016-03-29 & 00010 & 0.053 & 0.008 & 1.019\\
& & & 7& 2016-04-27 & 00012 & 0.052 & 0.013 & 0.843\\
& & & 8& 2016-08-26 & 00029 & 0.092 & 0.012 & 1.001\\
& & & 9& 2016-11-19 & 00099 & 0.067 & 0.009 & 1.010\\
& & &10& 2016-11-26 & 00053 & 0.063 & 0.008 & 1.157\\
& & &11& 2017-02-06 & 00025 & 0.111 & 0.011 & 1.054\\
& & &12& 2017-03-19 & 00010 & 0.103 & 0.011 & 0.983\\
& & &13& 2017-04-23 & 00011 & 0.077 & 0.011 & 0.941\\
\hline
NGC\,2068& 5h 46m 13.00s & +00:06:05.00 & 1& 2015-12-26 & 00052 & 0.116 & 0.013 & 0.957\\
& & & 2& 2016-01-16 & 00027 & 0.058 & 0.008 & 1.145\\
& & & 3& 2016-02-06 & 00015 & 0.046 & 0.011 & 1.086\\
& & & 4& 2016-02-29 & 00013 & 0.042 & 0.012 & 0.991\\
& & & 5& 2016-03-29 & 00011 & 0.056 & 0.010 & 1.010\\  
& & & 6& 2016-04-27 & 00013 & 0.052 & 0.016 & 0.774\\
& & & 7& 2016-08-27 & 00053 & 0.083 & 0.012 & 0.966\\
& & & 8& 2016-11-20 & 00088 & 0.093 & 0.012 & 0.935\\
& & & 9& 2016-11-26 & 00056 & 0.063 & 0.009 & 1.132\\
& & &10& 2017-02-06 & 00017 & 0.113 & 0.012 & 1.045\\
& & &11& 2017-03-19 & 00014 & 0.112 & 0.012 & 1.020\\ 
& & &12& 2017-04-21 & 00025 & 0.091 & 0.015 & 0.846\\
\hline
OMC\,23 & 5h 35m 31.00s & -05:00:38.00 & 1& 2015-12-26 & 00036 & 0.109 & 0.011 & 1.039\\
& & & 2& 2016-01-16 & 00019 & 0.057 & 0.009 & 1.013\\
& & & 3& 2016-02-06 & 00012 & 0.041 & 0.011 & 1.023\\
& & & 4& 2016-02-29 & 00011 & 0.043 & 0.012 & 0.917\\
& & & 5& 2016-03-25 & 00015 & 0.056 & 0.011 & 0.974\\
& & & 6& 2016-04-22 & 00011 & 0.050 & 0.011 & 0.966\\
& & & 7& 2016-08-26 & 00020 & 0.106 & 0.014 & 1.016\\
& & & 8& 2016-11-26 & 00052 & 0.062 & 0.008 & 1.125\\
& & & 9& 2017-02-06 & 00021 & 0.115 & 0.011 & 1.032\\
& & &10& 2017-03-18 & 00012 & 0.102 & 0.011 & 0.959\\
& & &11& 2017-04-21 & 00022 & 0.090 & 0.014 & 0.865\\
\hline
Oph& 16h 27m 05.00s & -24:32:37.00& 1& 2016-01-15 & 00084 & 0.067 & 0.012 & 0.913\\
Core& & & 2& 2016-02-05 & 00063 & 0.038 &  0.011 & 1.100\\  
& & & 3& 2016-02-26 & 00051 & 0.045 &  0.010 & 1.095\\ 
& & & 4& 2016-03-19 & 00065 & 0.044 &  0.011 & 1.073\\
& & & 5& 2016-04-17 & 00043 & 0.038 &  0.011 & 1.048\\
& & & 6& 2016-05-21 & 00034 & 0.076 &  0.016 & 0.933\\
& & & 7& 2016-08-26 & 00011 & 0.108 &  0.018 & 0.935\\
& & & 8& 2017-02-06 & 00083 & 0.110 &  0.013 & 0.998\\ 
& & & 9& 2017-03-20 & 00053 & 0.062 &  0.010 & 1.042\\ 
& & &10& 2017-04-19 & 00033 & 0.105 &  0.015 & 1.003\\
& & &11& 2017-05-18 & 00028 & 0.039 &  0.011 & 1.119\\
& & &12& 2017-06-16 & 00010 & 0.124 &  0.013 & 1.028\\
\hline
Serpens & 18h 29m 49.00s & +01:15:20.00 & 1& 2016-02-02 & 00054 & 0.091& 0.012 & 0.972\\
Main& & & 2& 2016-02-23 & 00050 & 0.050 & 0.011 & 1.032\\
& & & 3& 2016-03-17 & 00051 & 0.040 & 0.012 & 0.954\\
& & & 4& 2016-04-15 & 00046 & 0.040 & 0.010 & 1.118\\
& & & 5& 2016-05-21 & 00039 & 0.077 & 0.014 & 0.988\\
& & & 6& 2016-07-22 & 00023 & 0.096 & 0.012 & 0.989\\
& & & 7& 2016-08-27 & 00012 & 0.087 & 0.011 & 1.012\\
& & & 8& 2016-09-29 & 00012 & 0.094 & 0.012 & 0.963\\
& & & 9& 2017-02-22 & 00070 & 0.097 & 0.010 & 1.077\\
& & &10& 2017-03-20 & 00056 & 0.066 & 0.010 & 1.029\\
& & &11& 2017-04-03 & 00053 & 0.063 & 0.010 & 1.041\\
& & &12& 2017-04-17 & 00044 & 0.057 & 0.009 & 1.133\\
& & &13& 2017-05-05 & 00035 & 0.046 & 0.008 & 1.179\\
& & &14& 2017-05-19 & 00030 & 0.105 & 0.012 & 1.049\\
& & &15& 2017-06-02 & 00041 & 0.079 & 0.010 & 1.086\\
& & &16& 2017-06-16 & 00025 & 0.100 & 0.010 & 1.027\\
\hline
Serpens&  18h 29m 62.00s & -02:02:48.00& 1& 2016-02-02 & 00058 & 0.093 & 0.011 & 0.974\\
South& & & 2& 2016-02-23 & 00065 & 0.053 & 0.010 & 1.054\\
& & & 3& 2016-03-17 & 00052 & 0.042 & 0.013 & 0.911\\
& & & 4& 2016-04-15 & 00048 & 0.040 & 0.010 & 1.085\\
& & & 5& 2016-05-21 & 00044 & 0.074 & 0.013 & 0.983\\
& & & 6& 2016-07-21 & 00011 & 0.077 & 0.012 & 1.024\\
& & & 7& 2016-08-27 & 00017 & 0.092 & 0.012 & 0.979\\
& & & 8& 2016-09-29 & 00018 & 0.083 & 0.012 & 0.931\\
& & & 9& 2017-02-22 & 00081 & 0.099 & 0.009 & 1.165\\
& & &10& 2017-03-22 & 00094 & 0.068 & 0.010 & 1.085\\
& & &11& 2017-04-19 & 00036 & 0.109 & 0.015 & 1.028\\
& & &12& 2017-05-18 & 00037 & 0.042 & 0.010 & 1.098\\
& & &13& 2017-06-16 & 00021 & 0.104 & 0.011 & 1.007\cr
\enddata
\tablenotetext{a}{The optical depth at 225\,GHz.}
\tablenotetext{b}{The Flux calibration shown is the factor applied to the epoch after data reduction using the observatory-defined default calibration factor.}

\end{deluxetable}

For each of the eight observed regions, the individual calibrated epochs were stacked to produce a deep mean image.  The source finding routine JSA\_Catalogue {[found in Starlink's PICARD package \citep{sun265}]} was then applied to search for locations of peak brightness. {JSA\_Catalogue optimizes the user inputs of the FellWalker algorithm \citep{berry15} to identify compact, peaked emission and the associated, larger-scale structure. In this work, only the former is considered. Identified peaks must have a brightness at least 5 times the locally derived rms noise and an area of at least 81 square arcseconds (9 pixels). {These pixels must be adjacent and the peak width in both the horizontal and vertical dimensions must be at least 2 pixels.} Nearby peaks are merged into a single object if the flux between them does not drop by a level of at least 5 times the rms noise.} 

The resulting 1643 sub-mm peaks were then collated against master catalogues of disks (Class II) and protostars (Class 0/I) identified by the {\it Spitzer Space Telescope} \citep{megeath12,dunham15} and the {\it Herschel Space Observatory} \citep{stutz13}, with matches assumed for sources separated by less than 10\arcsec \citep[for details 
see][]{mairs17}.\footnote{One source in NGC\,2068 was originally classified as starless since it is not found in either the {\it Spitzer} catalogue of Orion YSOs \citep{megeath12} or the {\it Herschel} supplementary list of new protostars \citep{stutz13}. After being identified as a robust secular variable (see \S\ref{Sec:Slope} below), a literature search uncovered it as a PACS Bright Red Source (PBRS) identified by \citet{stutz13}. PBRS properties are consistent with those of young Class 0 protostars. The source is known as HOPS 373 \citep[see also][]{furlan16}.} Table\ \ref{Tab:SourceList} provides statistics on the numbers of sources of various type identified in each region as a function of limiting peak brightness. In the table we separate out the 151 sources brighter than 0.35 Jy\,bm$^{-1}$, for which the fractional  uncertainty in the individual measurements is less than $5\%$ (see Eqn\ \ref{Eqn:Fiducial} in \S \ref{Sec:Sigma}).

%chopped from body of text
\begin{deluxetable}{lcccccc}
\tablecolumns{7}
\tablewidth{0pc}
\tablecaption{Number of Peaks Extracted from Co-added Images \label{Tab:SourceList}}
\tablehead{
\colhead{Region} &
\colhead{All Peaks} & 
\colhead{Peaks}  &
\colhead{All Protostars} & 
\colhead{Protostars }  &
\colhead{All Disks} & 
\colhead{Disks} \\
\colhead{}&
\colhead{}&
\colhead{($>0.35\,$Jy\,bm$^{-1}$)}  &
\colhead{}&
\colhead{($>0.35\,$Jy\,bm$^{-1}$)}  &
\colhead{}&
\colhead{($>0.35\,$Jy\,bm$^{-1}$)} \cr
}
\startdata
IC\,348&         79&  3&  6&  3&  5& 0\\
NGC\,1333&      142& 18& 24& 11&  8& 0\\
NGC\,2024&      369& 11&  5&  2& 15& 1\\
NGC\,2068&      126& 19& 15& 10&  6& 0\\
Oph\,Core&      257& 14& 24&  2& 24& 4\\
OMC\,23&        276& 54& 15& 13& 40& 9\\
Serpens\,Main&   92& 12&  8&  7&  5& 0\\
Serpens\,South& 302& 20& 18&  3&  8& 0\\
\hline
Total&         1643&151&115& 51&111& 14\cr
\enddata
\end{deluxetable}

The source identification algorithm used in this paper is significantly different than that used by \citet{mairs17} to calibrate individual epochs. Mairs et al.\ use Gaussclumps \citep{stutzki90} to fit Gaussian profiles to peaks in individual region maps and then group Gaussians that share common coordinates across epochs. By fitting Gaussians, \citet{mairs17} are able to centroid sources in images that are not yet spatially aligned, a key component of their calibration analysis. In the present analysis we identify the peak brightness location in the stacked mean map and then use this fixed location to compare across individual epochs, making explicit use of the calibration and alignment achieved by \citet{mairs17}

\section{Stochasticity: Standard Deviation Analysis}
\label{Sec:Sigma}

The large number of sources (greater than 1500 overall and more than 150 with peak brightness larger than 0.35\,Jy\,bm$^{-1}$), the multiple epochs (typically twelve per region), and the consistent observing conditions and calibration procedures provide a rich, uniform data set to inspect. We therefore search for evidence of variability, either stochastic or secular, via statistical investigations. 

\begin{figure}[htb]
\center\includegraphics[scale=0.52]{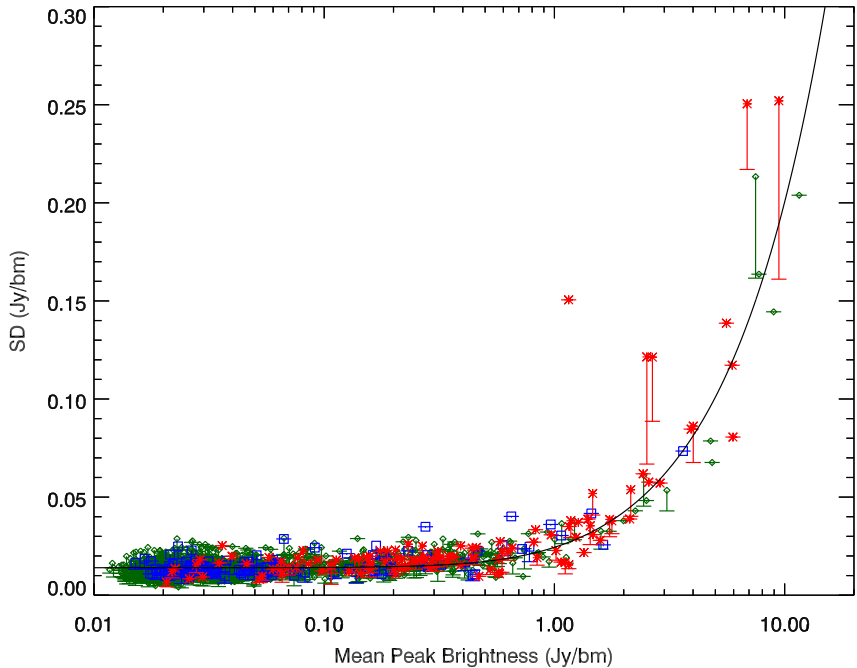}
\caption{Scatter plot of measured {mean peak brightness} versus measured peak brightness standard deviation for all sources in the JCMT Transient Survey. Individual sources are colored green (diamonds) if they are starless, blue (squares) if they are associated with known disks, and red (stars) if they are associated with known protostars. The solid line denotes the fiducial model,  $SD_{\rm fid}$, for the expected source brightness uncertainty as a function of brightness (see Eqn.\ \ref{Eqn:Fiducial}). The symbols show the measured peak brightness standard deviation when all epochs are used while the lower limits plot the result of sigma-clipping the measurement set (see text).}
\label{Fig:SD1}
\end{figure}

The first analysis that we perform is to measure the standard deviation of the brightness of each individual source, across all eight regions. For each source, $i$, the mean peak brightness in the stacked image, $f_m(i)$, is compared against the brightness at the same location in each individual epoch $n$, $f_n(i)$. The source variance, across all epochs, is then $v(i) = \sum \left[f_m(i) - f_n(i)\right]^2$ and the standard deviation is $SD(i) = \left[ v(i)/(n_e-1) \right]^{1/2}$ where $n_e$ is the number of epochs for which the region has been observed. The peak brightness standard deviation measurements for all sources are plotted against mean peak brightness in Figure\ \ref{Fig:SD1}, with colors denoting whether the source is associated with a known protostar (Class 0/I: red star) or a disk source (Class II: blue square). 

We expect that the uncertainty in the brightness of faint sources will be dominated by the relatively uniform noise in a given epoch, $\Delta f({\rm faint}) \sim 0.014\,$Jy\,bm$^{-1}$, while for bright sources the uncertainty in the relative calibration of the map will dominate, $\Delta f({\rm bright}) \sim 0.02\,f({\rm bright})$ \citep[see][]{mairs17b}. Thus, we present a fiducial standard deviation model, $SD_{\rm fid}$, where
\begin{equation}
SD_{\rm fid}(i) = \left[ \left(0.014 \right)^2 + \left(0.02 \times f_m(i)\right)^2 \right]^{1/2} {\rm Jy\, bm}^{-1}.
\label{Eqn:Fiducial}
\end{equation}

\begin{figure}[htb]
\center\includegraphics[scale=0.52]{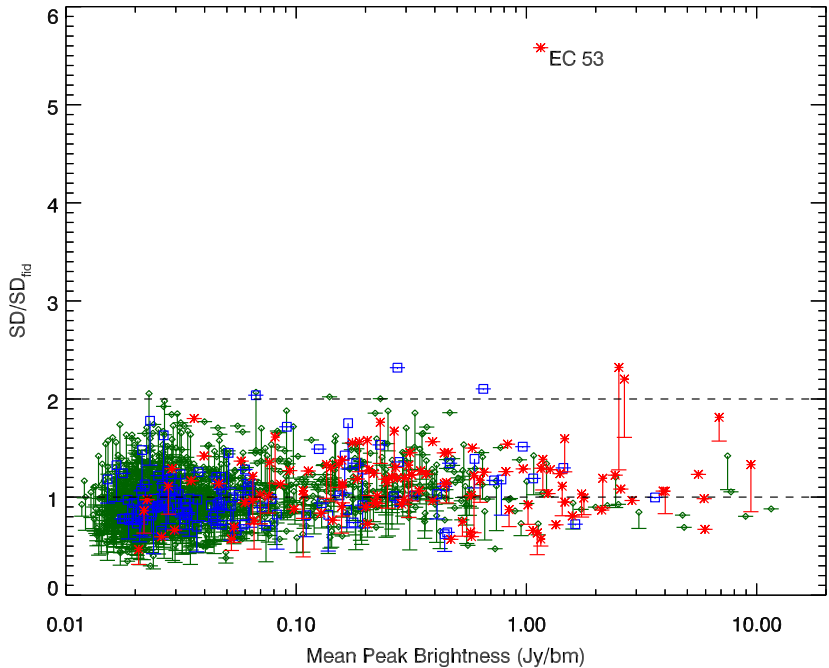}
\caption{Scatter plot of measured {mean peak brightness} versus measured peak brightness standard deviation divided by the fiducial model for all sources in the JCMT Transient Survey. Individual sources are colored green (diamonds) if they are starless, blue (squares) if they are associated with known disks, and red (stars) if they are associated with known protostars. The dashed lines indicate unity and twice the fiducial expectation. The symbols show the results when all epochs are used while the lower limits plot the results after sigma-clipping the measurement set (see text). Only one source stands out, EC\,53 \citep{yoo17}, a known variable protostar in Serpens Main.}
\label{Fig:SD2}
\end{figure}

The majority of the standard deviation measures in Figure\ \ref{Fig:SD1}  lie near the fiducial model, with some scatter.  To show this result more clearly, Figure\ \ref{Fig:SD2} plots the standard deviation in units of the fiducial model as a function of source brightness. Only one source stands out in this plot.  EC\,53, a protostar in Serpens Main with a peak flux of 1.15\,Jy\,bm$^{-1}$, was already identified as a sub-mm variable in our survey \citep{yoo17}, and was previously known as a Class I periodic variable at 2\,$\mu$m \citep{hodapp12}. EC\,53 has a measured uncertainty greater than five times the fiducial standard deviation. No other sources have uncertainties greater than 2.5 times the fiducial standard deviation. Four of the brighter sources with $f_m > 0.2\,$Jy\,bm$^{-1}$, two each of protostars and disks, have marginal enhanced uncertainties, appearing slightly beyond the bulk of the ensemble in Figure \ref{Fig:SD2}. We investigate these outliers, along with EC\,53, in Section\ \ref{Sec:Individual} and present all five sources in Table\ \ref{Tab:SD}.

% Figure 1

% Figure 2

%chopped from body of text
\begin{deluxetable}{lrrcclcr}
\tablecolumns{8}
\tablewidth{0pc}
\tablecaption{Potential Stochastic Variable Sub-mm Sources \label{Tab:SD}}
\tablehead{
\colhead{Region} &
\colhead{RA} & 
\colhead{Dec}  &
\colhead{Peak Brightness} & 
\colhead{(SD/SD$_{\rm fid})$\tablenotemark{a}}  &
\colhead{Source} & 
\colhead{Outlier Events\tablenotemark{b}}&
\colhead{Identification} \\
\colhead{}&
\colhead{}&
\colhead{}&
\colhead{(Jy\,bm$^{-1}$)}&
\colhead{}&
\colhead{Type}&
\colhead{}&
\colhead{}\cr
}
\startdata
Serpens\, Main& 18:29:51.2& +01:16:38& 1.15& 5.6& Protostar& &EC\,53\\
OMC\,23& 05:35:22.4& -05:01:11& 2.55& 2.3& Protostar& 
5.2\,$\sigma$; Epoch 3& HOPS\,88\\
OMC\,23& 05:35:25.2& -05:19:14&0.28&2.3& Disk& & V1017 Ori\\
OMC\,23& 05:35:27.4& -05:09:29&2.69&2.2& Protostar& 
6.5\,$\sigma$; Epoch 10& HOPS\,370\\
Oph\,Core& 16:26:24.4& -24:16:16&0.65&2.1& Disk& & Oph 162624\cr
\enddata
\tablenotetext{a}{The fiducial flux density standard deviation model is described in Section 3.}
\tablenotetext{b}{Shown are individual events greater than five times the measured standard deviation of the source.}
\end{deluxetable}

An important concern when dealing with small number statistics is the influence of outlier data points in our $\sim 12$ epochs. The occasional extreme data point will skew the measured uncertainty for the affected source unless the data are sigma-clipped, whereby data points well away from the mean are removed from the measurement set before statistical analysis. Additionally, given the large number of sources investigated, these extreme measurements happen relatively frequently within the sample even if rarely for a given object. To complicate this situation further, one of the primary goals of the JCMT Transient Survey is to uncover rare transient phenomena which will be mimicked by purely statistical large random deviations. Thus, much care is required to separate true stochastic behavior from random noise statistical outliers.

A detailed, robust investigation into the statistics of outlier points is beyond the scope of this initial results paper and will be more easily investigated as the number of epochs increase substantially, by a factor $\sim 2$, over the lifetime of the JCMT Transient Survey. Here we perform only a simple test for outlier brightness data points. For all epochs of each source, we calculate the deviation between the brightness measured versus the mean brightness over all epochs in units of the measured standard deviation. We then remove (sigma-clip) those epochs that vary by more than two standard deviations and recompute both the mean and the standard deviation for the source. A total of 675 individual measurements $\left[ f_n(i) \right]$ are removed by this process, representing about 5\% of the entire measurement set as expected statistically for two-sigma clipping of a normal distribution.  In Figures\ \ref{Fig:SD1} and \ref{Fig:SD2}, the downward error bars plot the change in the measured standard deviation for those sources where such outlier brightness measurements are removed. Note that the two protostars with initial standard deviation measures somewhat larger than the ensemble fiducial significantly regress toward the standard deviation model when their outlier points, one per source, are removed (see Figure\ \ref{Fig:SD2}, Table\ \ref{Tab:SD}, and \S \ref{Sec:Individual}). EC\,53 and the two disk sources noted above are not influenced by the sigma-clipping process, {i.e., no points are removed.}

We have further analyzed the statistics of these 675 outlier brightness measurements and note that there is a significant skew in the distribution toward measurements higher than the mean, as opposed to outlier measurements for which the peak brightness was observed to decrease. While a possibly intriguing physical result, we caution that at least some of this may be due to data reduction artifacts, such as noise spikes in the original data set that have not been entirely removed before the map-making process. To be confident that the observed skew is due to a physical phenomenon associated with the observed sub-mm sources will require a more detailed investigation of where and when in each mapped region these outlier points are observed, including consideration of locations well away from known peaks. This investigation will be much easier to accomplish with the larger data set available at the end of the JCMT Transient Survey. We thus leave a full discussion of stochastic and transient phenomena to the survey summary paper and here only note that, of the 675 outlier data points uncovered, only seven are found to have offsets greater than than 5 times the source standard deviation. Based on normal statistics, the anticipated number of false-positives with such large offsets is $\ll 1$ over the entire sample. Two of these extreme outliers are identified with sources brighter than 0.1 Jy\,bm$^{-1}$  - the two protostars discussed above and in \S\ \ref{Sec:Individual}  (see also Table\ \ref{Tab:SD}).

In total, out of 1643 independent peaks analyzed, only EC\,53 is found to have a measured brightness standard deviation significantly larger than the expected fiducial.  Thus, less than 0.1\% of the ensemble peaks are found to be stochastically varying above the uncertainty in the survey measurements and less than 1\% of the protostellar sub-sample show such variability. None of the $> 1400$ peaks that are unassociated with known protostars show significant evidence for stochastic variability. 

\section{Secularity:  Linear Variability Analysis}
\label{Sec:Slope}
The results of the previous section confirm that, at the level of calibration which we are able to achieve with the JCMT Transient Survey and the time range of observing, there is only one source at present that is undergoing very large variations in the peak brightness measurement at sub-mm wavelengths \citep[EC 53; see also][]{yoo17,mairs17b}.
 The remaining sources all show scatter about the expected fiducial standard deviation measure with only a hint that a few of these may be variable.  Nevertheless, it is possible to investigate the range of (linear) secular brightness variations allowed by the measured brightnesses across all epochs and to search for statistically robust secular detections. Further, by comparing against the same brightness observations randomized in time-order, we can also place limits on the degree of secular variation lurking within the ensemble. Given that we are interested in constraining the range of possible secular variability, this analysis includes those sources with a fiducial brightness standard deviation (Eq.\ \ref{Eqn:Fiducial}) of $<5$\%, which corresponds to 
sources brighter than  $0.35\,$Jy\,bm$^{-1}$. Furthermore, since we are primarily interested in the statistical properties of the ensemble, we remove from the sample the strong detection, EC\,53. Thus,  as shown in Table\ \ref{Tab:SourceList}, we are left with a sample of 150 peaks, of which 50 are associated with known protostars (Class 0/I) and 14 are associated with disks. 

We begin by computing a least squares linear fit to the source brightness across all measured epochs using the IDL routine {\tt linfit}, which reproduces the {\tt fit} and {\tt gammq} routines described by \citet{press89}.
In this analysis we do not sigma-clip (see discussion in Section\ \ref{Sec:Sigma}) and we further assume that all measurements for a given source have the same uncertainty\footnote{The observing strategy for the JCMT Transient Survey ensures that each epoch has roughly the same sensitivity (see \citet{herczeg17} and \citet{mairs17}.}. Thus, for each source, $i$, we derive a model fit, $f_l(i,t)$, which is linear over time, $t$, with two derived parameters: initial flux, $f_0(i)$ at time $t_0$, the time of the first epoch, and slope, $S(i)$, measured in fractional brightness change over a year;
\begin{equation}
f_l(i,t) =  f_0(i)\,\left(1 + S(i)\,\left[ t - t_0) \right] \right).
\label{Eqn:Linear}
\end{equation}
Furthermore, in order to measure the relevance of any slopes departing from flat ($S=0$; no change in brightness over time) we also compute the  uncertainty in the slope, $\Delta S$, using the same IDL routine.\footnote{The derivation of $\Delta S$ includes both the uncertainty in the measurement of $S$ and the uncertainty in the measurement of $f_0$ \citep{press89}.} We then perform standard statistical analyses on the ensemble to determine the mean, standard deviation, and skewness of the set of measured slopes. These mean and standard deviation statistical measures are presented in Table \ref{Tab:Stats}. In all cases the skewness is found to lie statistically close to zero (that is the distribution is well enough described by a Gaussian) and therefore is not included in the table.

%chopped from body of text
\begin{deluxetable}{lrrrrrrrrr}
\tablecolumns{10}
\tablewidth{0pc}
\tablecaption{Secular Variability Analysis: Histogram Widths  \label{Tab:Stats}}
\tablehead{
\colhead{Ensemble} &
\multicolumn{4}{c}{Time-Ordered}&\colhead{}&\multicolumn{4}{c}{Randomized}\cr
\multicolumn{10}{c}{}\cr
\colhead{}&
\colhead{$\overline{S}$\tablenotemark{a}}&
\colhead{$\sigma_{S}$\tablenotemark{b}} &
\colhead{$\overline{S/\Delta S}$\tablenotemark{c}}&
\colhead{$\sigma_{S/\Delta S}$\tablenotemark{d}} &
\colhead{\ \ \ \ \ \ } &
\colhead{$\overline{S}$\tablenotemark{a}}&
\colhead{$\sigma_{S}$\tablenotemark{b}} & 
\colhead{$\overline{S/\Delta S}$\tablenotemark{c}}&
\colhead{ $\sigma_{S/\Delta S}$\tablenotemark{d}}\cr
}
\startdata
All Bright Sources    & -0.006& 0.023& -0.27& 1.25& &0.000& 0.022& -0.02& 1.11\\
Clipped\tablenotemark{e} Bright Sources& -0.007& 0.022& -0.27& 1.06& &0.000& 0.022&  0.02& 1.04\\
\hline
Protostars Sample        & -0.005& 0.025&  -0.25& 1.61& &0.000& 0.020& -0.04& 1.12\\
Clipped\tablenotemark{e} Protostars Sample& -0.006& 0.020&  -0.27& 1.11& &0.000& 0.020& -0.04& 1.04\cr 
\enddata
\tablenotetext{a}{Measured mean value of the ensemble of slope measurements.}
\tablenotetext{b}{Measured width of the ensemble of slope measurements.}
\tablenotetext{c}{Measured mean value of the ensemble of slope divided by slope uncertainty measurements.}
\tablenotetext{d}{Measured width of the ensemble of slope divided by slope uncertainty measurements.}
\tablenotetext{e}{For the clipped sample we remove all sources with $|S/\Delta S| >3$ (see Section 4).}
\end{deluxetable}

% Figure 3
\begin{figure}[htb]
\center{
\includegraphics[scale=0.30]{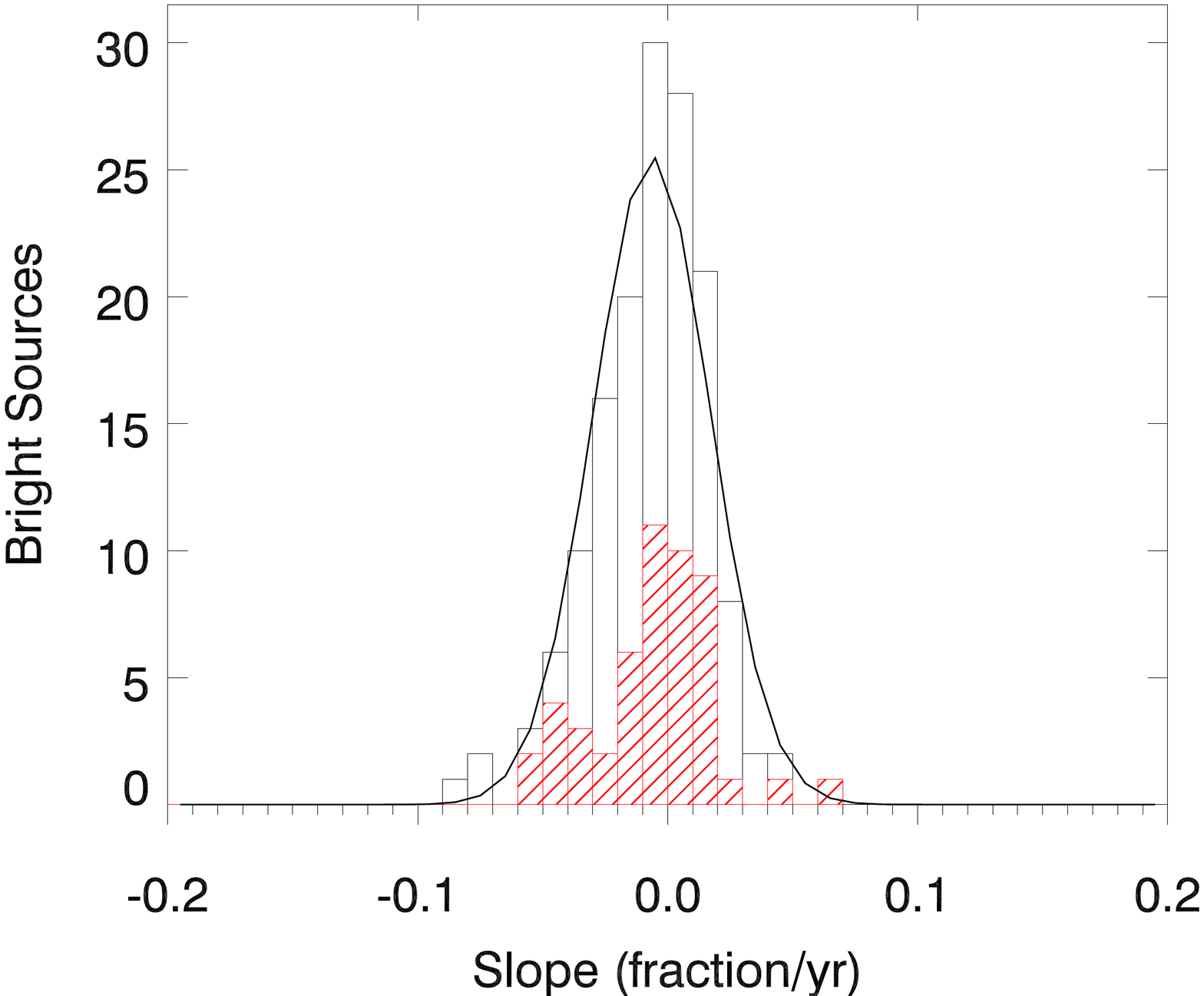}
\includegraphics[scale=0.30]{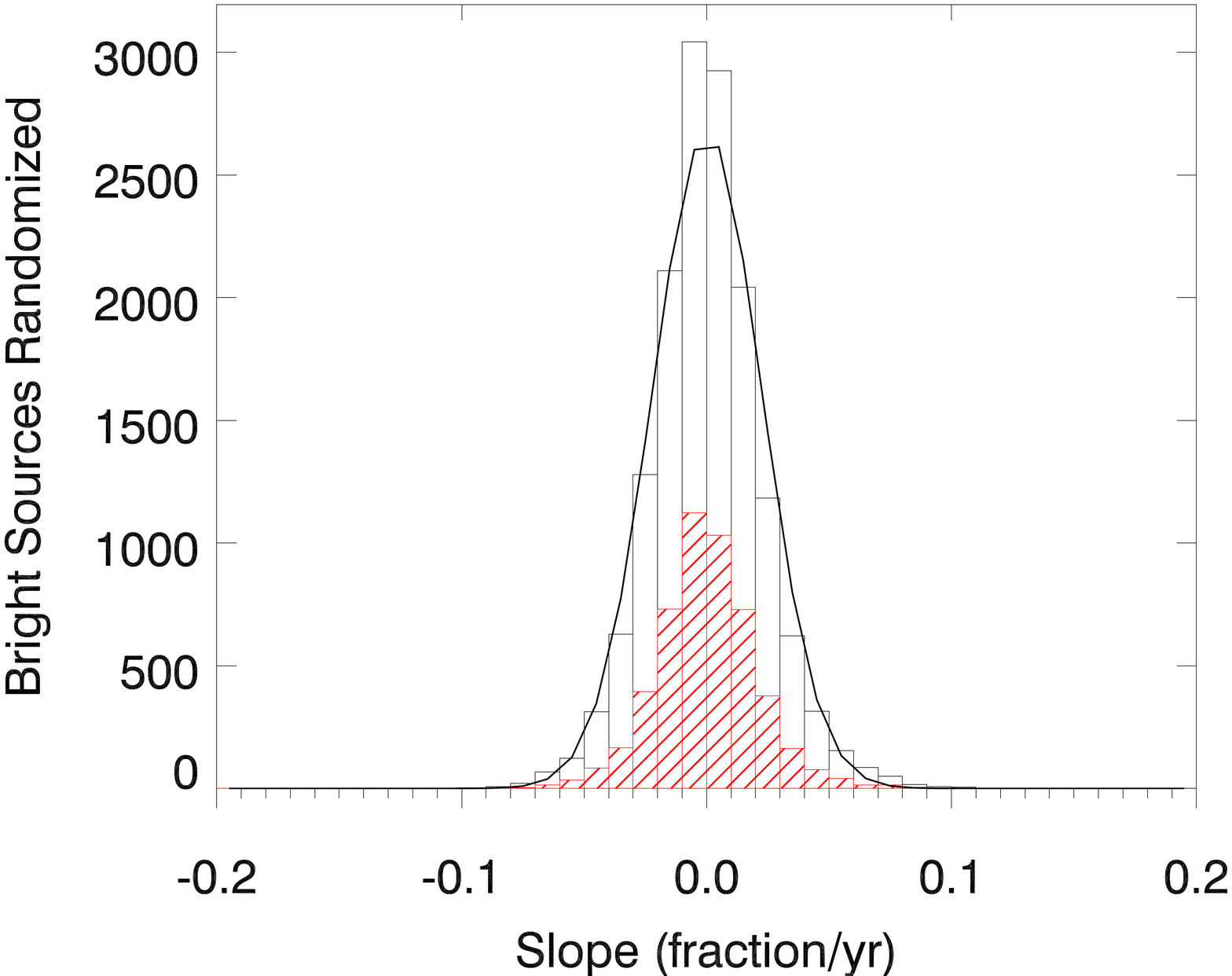}
}
\caption{Histograms of the fractional peak brightness change over a year for the ensemble of 150 bright sources in the JCMT Transient Survey. The 50 sources known to be associated with protostars are shown over-plotted in red. Left Panel: Results from least square fitting to all epochs for each source.  Right Panel: Results from least square fitting to all epochs for each source after randomly time-ordering the peak flux measurements 100 times. The parameters of the statistical fits are provided in Table \ref{Tab:Stats}. Note that the visible outlier measurements in the left panel are {\it not} significant (see text).}
\label{Fig:Slopehist}
\end{figure}

The left panel of Figure\ \ref{Fig:Slopehist} plots the histogram of the ensemble of measured slopes.  The distribution is well fit by a Gaussian profile centered near zero with $\sigma_S = 0.023$ (where $S= \pm 0.023$ corresponds to an increase or decrease in flux of 2.3\% over a year; see Table \ref{Tab:Stats}), implying that a majority of the sub-mm peaks show little evidence for strong secular variability over year timescales.  The protostar subset (red, hashed) follows the same overall distribution, with $\sigma_{S_p} = 0.025$. While there are too few disks to obtain reliable statistical measures, the distribution shows no obvious disagreement with the full ensemble (not shown in the figure). 

There are a few outlier slope measurements seen in the histogram (4 with slope magnitudes larger than 0.06). To test if these are statistically significant, for each source we keep the same epoch dates and peak brightness measurements but randomize the order in which the measurements are observed. In this way, the mean brightness and standard deviation for each source remains the same but any true secular variability will be removed. We are thus able to estimate the importance of {\it false positive} secular variability results appearing in the wings of the distribution. The right panel of Figure\ \ref{Fig:Slopehist} plots the randomized histogram of the ensemble of measured slopes after 100 independent randomizations for each source and the parameters of the statistical fits are shown in Table \ref{Tab:Stats}. Again, a Gaussian profile fits the histogram very well, with an almost identical width to the ordered observations, $\sigma_S({\rm random}) = 0.022$ [$\sigma_{S_p}({\rm random)} = 0.020$, slightly lower than the width of the time-ordered protostellar sample]. The fraction of randomized slope magnitudes larger than 0.06 (248 out of 15000, or $1.6\%$ of the sample) is similar to the fraction of outlier sources in the left hand plot (four outliers, $2.7\%$ of the sample), arguing that these outliers are {\it not} statistically significant.

\begin{figure}[htb]
\center{
\includegraphics[scale=0.3]{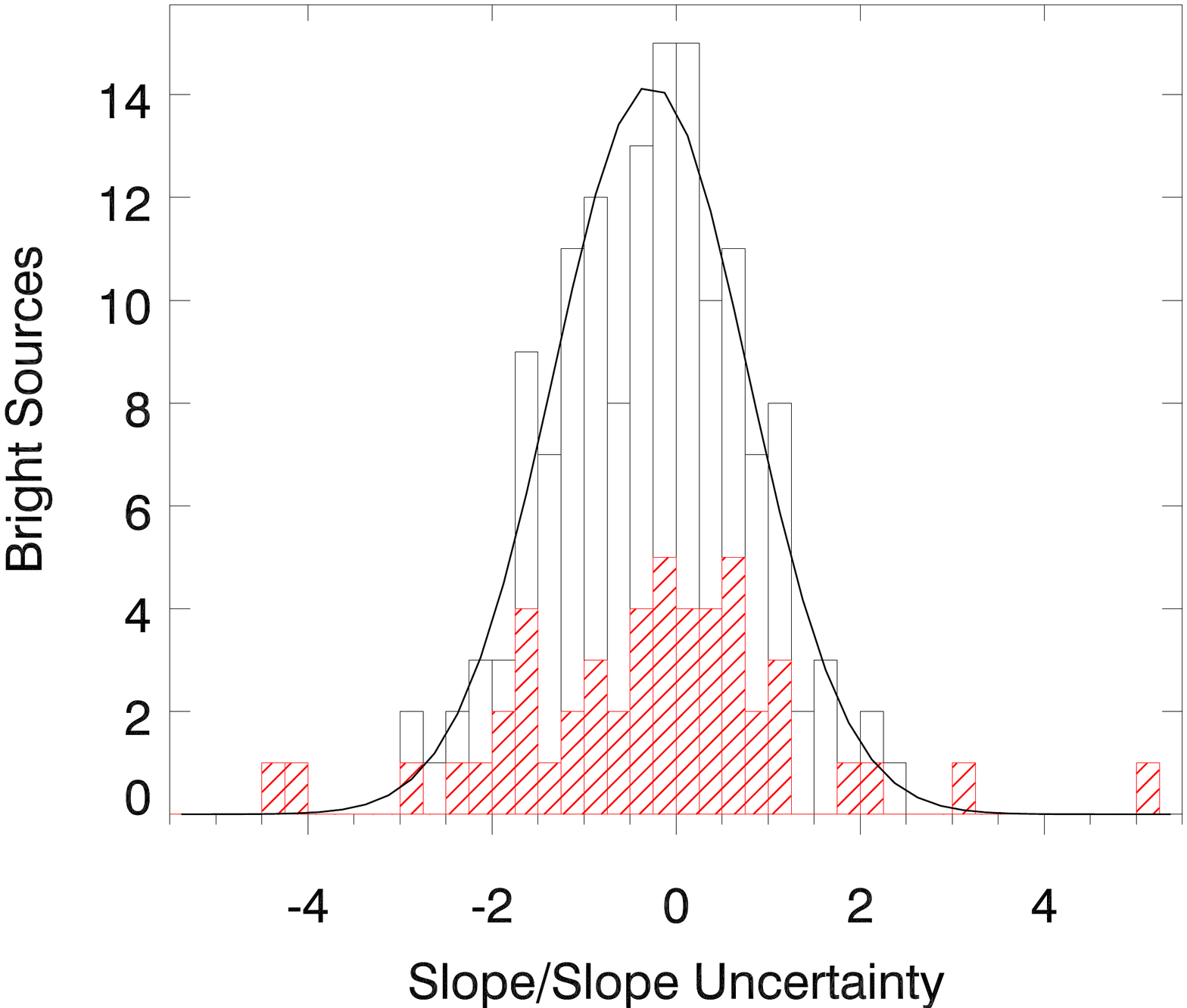}
\includegraphics[scale=0.3]{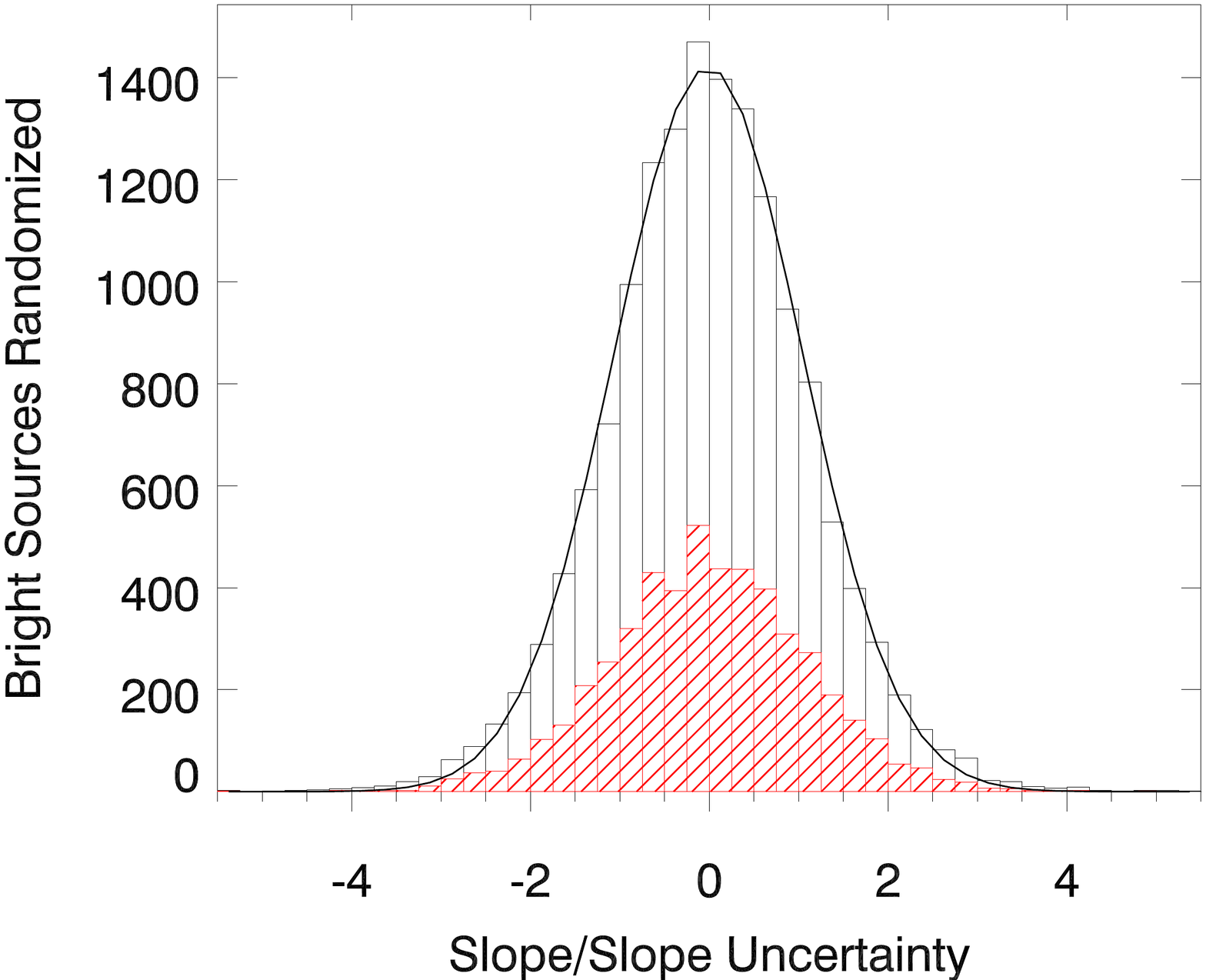}
}
\caption{Histograms of the fractional peak brightness change over a year divided by the uncertainty in the fractional peak brightness change over a year for the ensemble of 150 bright sources in the JCMT Transient Survey. The 50 sources known to be associated with protostars are over-plotted in red.  Left Panel: Results from least square fitting to all epochs for each source.  {Note that the right-most source in the left panel is SMM\,10 in Serpens Main ($S / \Delta S = 5.1$). The source EC\,53 is off the plot to the right ($S / \Delta S = 7.9$).} Right Panel: Results from least square fitting to all epochs for each source after randomly time-ordering the peak flux measurements 100 times. The parameters of the Gaussian fits are provided in Table \ref{Tab:Stats}. Note that the outlier measurements in the left panel are significant (see Table\ \ref{Tab:Secular} and text).}
\label{Fig:Slopefithist}
\end{figure}

Despite the lack of a clear signal in the measured slope histograms above, it is still possible that a subset of the bright sources may have statistically significant secular variation. We thus identify an additional goodness of fit metric for the slope measurements by taking the measured uncertainty in the slope and searching for those sources with $| S/\Delta S | \gg 1$.  Again, we determine the statistical properties of the ensemble and use the randomized data set to test the likelihood of this metric revealing otherwise hidden secular variables (see Table \ref{Tab:Stats}).  Figure\ \ref{Fig:Slopefithist} shows the distribution of $S/\Delta S$ for both the ordered measurements and 100 randomizations of each source. In both cases the histogram is well fit with $\sigma_{(S/\Delta S)} \sim 1$, as expected for a distribution dominated by uncertainty. More importantly, however, there exists a statistically relevant set of outliers in the ordered histogram, with $| S/\Delta S | > 3$.  We present these four sources, along with EC\,53, in Table\ \ref{Tab:Secular}, where we also compute the false-positive expectation value derived from the randomized distribution. All five sources are protostellar (Class 0/I), yielding an observed protostellar secular variability rate of 10\%. These sources are discussed individually in Section\ \ref{Sec:Individual} and the significance of secular detections only within the protostellar sample is presented in Section\ \ref{Sec:Discussion}. Recognizing that these outlier sources might skew the statistical analyses, we also provide results in Table \ref{Tab:Stats} for the ensemble statistics after sigma-clipping the sources with $|S/\Delta S| > 3$.

% Figure 4

The outlier points in Figure\ \ref{Fig:Slopefithist} appear to be true deviants and not just an extension of a smooth distribution, although the limited statistics makes this challenging to quantify (see Section\ \ref{Sec:Slope:Toy} below for additional comment). Furthermore, consideration of the light curves for these select objects, presented in Section\ \ref{Sec:Individual}, suggests that the assumption of a purely linear secular variation is a significant oversimplification and that more complex analyses might uncover additional detail once the survey is complete. For the present, the limited number of epochs and the limited signal to noise of the measurements versus the strength of any underlying secular variation, make this linear investigation the appropriate first analysis.

% Figure 5
\begin{figure}[htb]
\center{
\includegraphics[scale=0.5]{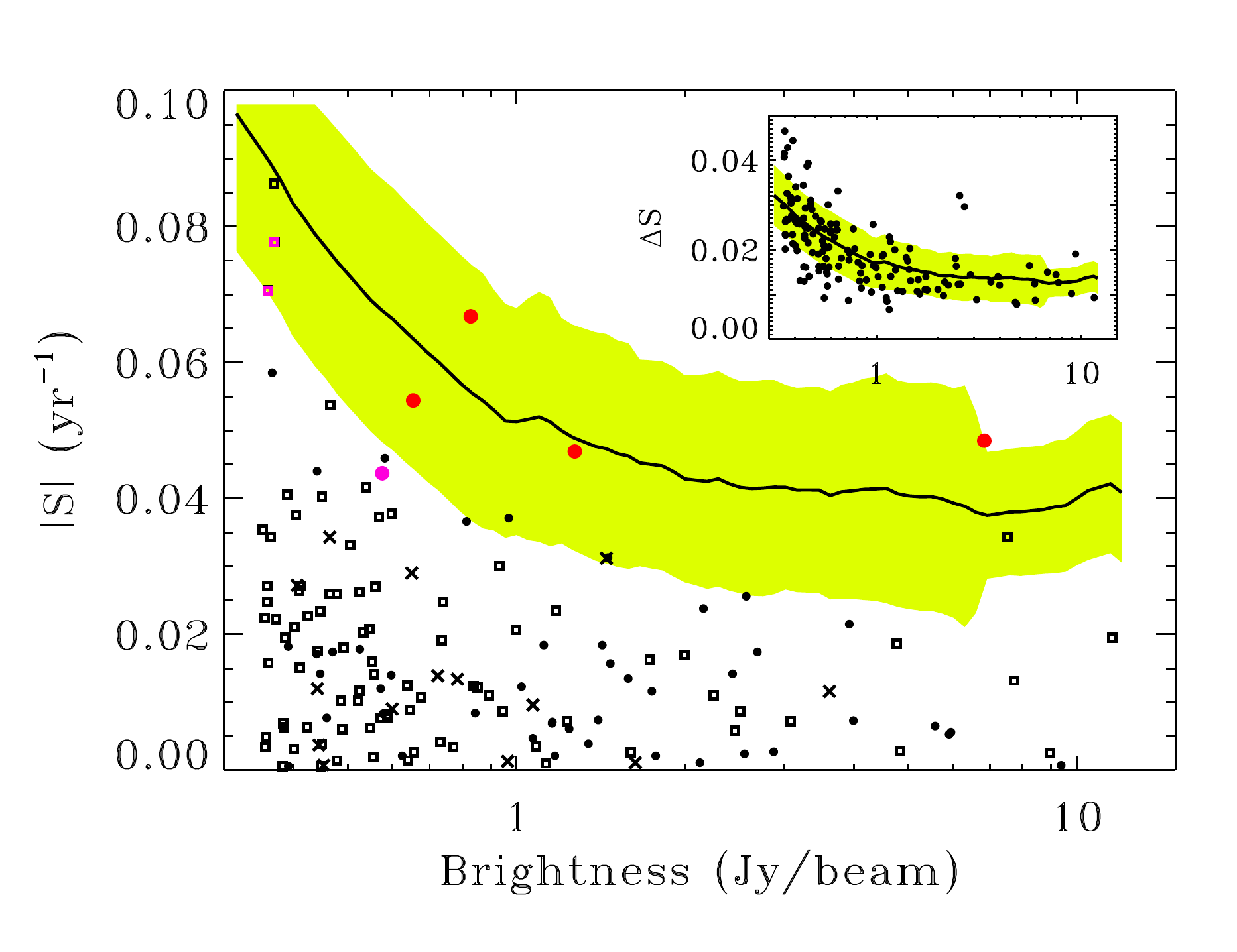}
}
\caption{
{
The distribution of the absolute value of the best-fit slope versus brightness (main plot), with an inset showing the uncertainty in slope, which increases to the faint end of our sample.  At any given brightness, a wide range in uncertainties leads to large differences in the sensitivity of brightness changes.  Filled circles denote known protostars, crosses denote disk sources, and squares represent starless cores. Symbols in red have $| S/\Delta S | > 3$ and symbols in purple have $2.5 < | S/\Delta S | <3$.  The solid line shows a smoothed (0.4 dex in brightness) running average of the values that would yield a $3\sigma$ detection for a typical source (the inset shows the $1\sigma$ detection limit), with a standard deviation in the uncertainty displayed as the shaded yellow region.  Some significant detections have slopes that are below the shaded region because the errors on those points are especially low, while other non-detections have slopes above the shaded region because of larger errors.}
}
\label{Fig:slopes}
\end{figure}

Figure \ref{Fig:slopes} shows the range in sensitivity to secular brightness changes for our sample as a function of source mean brightness.  The typical uncertainty in the slope, $\Delta S$, is independent of source brightness for sources brighter than 1 Jy beam$^{-1}$ (see inset) and increases linearly for fainter sources.  The distribution of uncertainties shows some scatter, since they depend on the individual source and the associated light curve.  
Robust detections of secular variability require $|S/\Delta S| > 3$ and are shown in red in the main figure.  
There is no evidence in the figure that secular variables are biased toward either brighter or fainter sub-mm sources.  However, given the individuality of the uncertainty measures, some sources are considered non-detections despite having steeper slopes than similar sources that are considered detections.

\begin{deluxetable}{lrrccclcr}
\tablecolumns{9}
\tablewidth{0pc}
\tablecaption{Potential Secular Variable Sub-mm Sources \label{Tab:Secular}}
\tablehead{
\colhead{Region} &
\colhead{RA} & 
\colhead{Dec}  &
\colhead{$f_0$\tablenotemark{a}} & 
\colhead{$S$\tablenotemark{a}}  &
\colhead{$S/\Delta S$\tablenotemark{a}}&
\colhead{Source} & 
\colhead{False Positive}&
\colhead{Identification} \\
\colhead{}&
\colhead{}&
\colhead{}&
\colhead{(Jy\,bm$^{-1}$)}&
\colhead{(yr$^{-1}$)}&
\colhead{}&
\colhead{Type}&
\colhead{Expectation\tablenotemark{b}}&
\colhead{}\cr
}
\startdata
Serpens\,Main& 18:29:51.2& +01:16:38& 0.95& 0.28& 7.9& Protostar& $\ll 0.01$&EC53\\
Serpens\,Main& 18:29:52.0& +01:15:50& 0.79& 0.07& 5.1& Protostar& 0.01& SMM\,10\\
NGC\,2068& 05:46:31.0&   -00:02:32& 1.31& -0.05& 4.3& Protostar& 0.08& HOPS\,373\\
Serpens\,South& 18:29:37.8& -01:51:03& 0.68& -0.05& 4.1& Protostar& 0.13& IRAS 18270-0153\\
Serpens\,Main & 18:29:49.8& +01:15:20&  6.6&  0.05& 3.2& Protostar& 1.33\tablenotemark{c}& SMM\,1\cr
\enddata
\tablenotetext{a}{Fit parameters to the linear model described in Section 4.}
\tablenotetext{b}{Number of false positives expected in a sample of 150 sources. If only considering the protostellar sample then the\\ false positive rate drops by a factor of three.}
\tablenotetext{c}{The measured slope versus measured slope uncertainty for this source is not extreme and thus the {\it false-positive}\\ expectation value provided is large. The source, however, is atypical and therefore it is unclear whether the ensemble\\ approach is applicable.  This is the brightest source in Serpens Main, among
the brightest five sources in the entire\\ ensemble, and was found to vary in brightness between the GBS and Transient Surveys \citep{mairs17b}.}
\end{deluxetable}

\subsection{Secular Variability Toy Model}
\label{Sec:Slope:Toy}
Along with the individual outlier secular variables found in the preceding section, there may well be additional secular signal buried within the ensemble results presented in Figures\ \ref{Fig:Slopehist}, \ref{Fig:Slopefithist}, and Table\ \ref{Tab:Stats}. To better understand the degree of secular variability allowed by the JCMT Transient Survey observations to date, and to predict the level of  improvement expected over the lifetime of the survey, we design a simple toy source model and repeat the analysis from \S \ref{Sec:Slope}.

For simplicity, we assume an ensemble of 10000 sources, each varying linearly in time with a slope, $S_t$, defining the fractional brightness change over a year, pulled from a normal distribution, $\sigma_{S_t}$. We further assume that these model sources are observed twelve times over 1.2 years with a regular cadence, approximately mimicking the available JCMT Transient Survey observations to date. Finally, we assume that the fractional uncertainty for each measurement for each source is $\Delta f/f = 0.025$, the typical uncertainty of our observed sample. In the absence of this measurement uncertainty, the slope returned by the linear fit to each model source (Eqn.\ \ref{Eqn:Linear}) would match exactly the input value; $S = S_t$ (and $\sigma_{S} = \sigma_{S_t}$).

In Table\ \ref{Tab:Toy} we present the statistical results of a linear least square analysis for the toy model ensemble, both prior to and after time-randomization, for $\sigma_{S_t} = 0.005$, $0.010$, and $0.020$. These results can be compared directly against the results from \S \ref{Sec:Slope} and Table \ref{Tab:Stats}.  As can be seen in the table, for a large value of $\sigma_{S_t}$ there is a significant difference in the width of the measured slope histogram between the original time-ordered data and the randomized data, with the original time-ordered data distribution being significantly broader than that of the randomized data. In all three model cases, the width of the measured slope for the randomized data remains similar to both the time-ordered and randomized data obtained by the JCMT Transient Survey to date, suggesting that the uncertainty assumptions of the toy model are reasonable. As the magnitude of $\sigma_{S_t}$ decreases, the width of the measured slope distribution for the time-ordered data becomes narrower. The trends seen here are entirely as expected since the strength of the input slope within the toy model is proportional to $\sigma_{S_t}$ and becomes directly measurable in the time-ordered data as the secular signal becomes greater than the uncertainty in the individual measurements. For the randomized data, the input slope is significantly disrupted and thus there is little signal remaining to influence the histogram. The observed trend is similar for the histograms of measured slope in units of the measured slope uncertainty. For large values of $\sigma_{S_t}$ this width becomes significantly larger than unity for the time-ordered observations while the randomized distribution remains close to 1.1 (which is also the value obtained by the Transient Survey ensemble).

%chopped from body of text
\begin{deluxetable}{lrrrrrrrrr}
\tablecolumns{10}
\tablewidth{0pc}
\tablecaption{Secular Variability Analysis: Toy Model Histogram Widths \label{Tab:Toy}}
\tablehead{
\colhead{Toy Model} &
\multicolumn{4}{c}{Results for Present Survey}&
 &
\multicolumn{4}{c}{Results after Full Survey}\cr
\colhead{}&\multicolumn{2}{c}{Time-Ordered}&\multicolumn{2}{c}{Randomized}&
\colhead{}&\multicolumn{2}{c}{Time-Ordered}&\multicolumn{2}{c}{Randomized}\cr
\colhead{}&
\colhead{$\sigma_{S}$\tablenotemark{a}} & 
\colhead{ $\sigma_{S/\Delta S}$\tablenotemark{b}} &
\colhead{$\sigma_{S}$\tablenotemark{a}} & 
\colhead{ $\sigma_{S/\Delta S}$\tablenotemark{b}}&
\colhead{}&
\colhead{$\sigma_{S}$\tablenotemark{a}} & 
\colhead{ $\sigma_{S/\Delta S}$\tablenotemark{b}} &
\colhead{$\sigma_{S}$\tablenotemark{a}} & 
\colhead{ $\sigma_{S/\Delta S}$\tablenotemark{b}}\cr
}
\startdata
$\sigma_{S_t} = 0.020$& 0.028& 1.62& 0.020& 1.12& &0.021& 4.23& 0.006& 1.04\\ 
$\sigma_{S_t} = 0.010$& 0.022& 1.27& 0.019& 1.11& &0.011& 2.27& 0.005& 1.04\\
$\sigma_{S_t} = 0.005$& 0.020& 1.15& 0.019& 1.11& &0.007& 1.47& 0.005& 1.03\cr
\enddata
\tablenotetext{a}{Standard deviation of $S$ measured for all members of the ensemble (see text).}
\tablenotetext{b}{Standard deviation of $S/S_e$ measured for all members of the ensemble (see text).}
\end{deluxetable}

The toy model can be extended in order to predict the detection power of the JCMT Transient Survey after its full three years of observation. By increasing the number of observations to 30 and the time period to three years, we show in Table\ \ref{Tab:Toy} that the difference in ensemble histogram widths between the time-ordered and random observations becomes much stronger for each value of  $\sigma_{S_t}$.  We thus conclude that at present we can rule out a Gaussian distribution of secular variations linear in time with $\sigma_{S_t} > 0.02$ but that a distribution with $\sigma_{S_t} < 0.01$ would be undetectable today. After the full three years of the survey, we expect to either measure $\sigma_{S_t}$ or rule out distributions with $\sigma_{S_t} > 0.005$.

In each of these toy models, the normal distribution of linear slopes prevents there being a significant population of sources with large brightness variations, such as those observed by the JCMT Transient Survey (Table\ \ref{Tab:Secular}).  This suggests that the five observed sources are indeed rare, statistical outliers. It should be recognized, however, that the toy models are meant to be representative rather than conclusive and that there remain many alternate secular variation distributions, for example modified power-laws, that might be able to fit the noise-dominated ensemble measurements, similarly to the imposed normal distribution, while also allowing for a few rare outliers. Once the full survey is complete it will be appropriate to consider these more complex distributions.

\newpage
\section{Individual Sources of Interest}
\label{Sec:Individual}

In the above investigation, we searched for stochastic and secular variables within the entire JCMT Transient Survey. In this section we discuss individually each candidate variable identified in this survey, along with several variables identified in \citet{mairs17b} and the sub-mm variable HOPS 383, identified by \citet{safron15}.  Where possible we compare those sources with robust secular fits with the brightness change measurements obtained by \citet{mairs17b}. Mairs et al.\ collated mean source brightnesses from the Transient Survey against archival JCMT Gould Belt Survey data \citep{ward-thompson07} to investigate variability over longer, 2-4 year timescales, uncovering five strong candidates. As discussed in Section \ref{Sec:Discussion} and shown in Table\ \ref{Tab:Sources}, there is significant overlap between the \citet{mairs17b} sample and the results from this paper. \citet{mairs17b} did not specifically consider stochastic variability and the analysis technique used down-weighted sources with large variations within either or both data Surveys. Thus we are not able to compare directly the possible stochastic variables found here with that paper. At the end of this section we present light curves for three sources that are not found to vary in the present analysis - the brightest sources observed in the OMC\,2/3 and Ophiuchus Core regions and the second brightest source in Serpens Main.

% Figures
\begin{figure}[htb]
\center{
\includegraphics[scale=0.25]{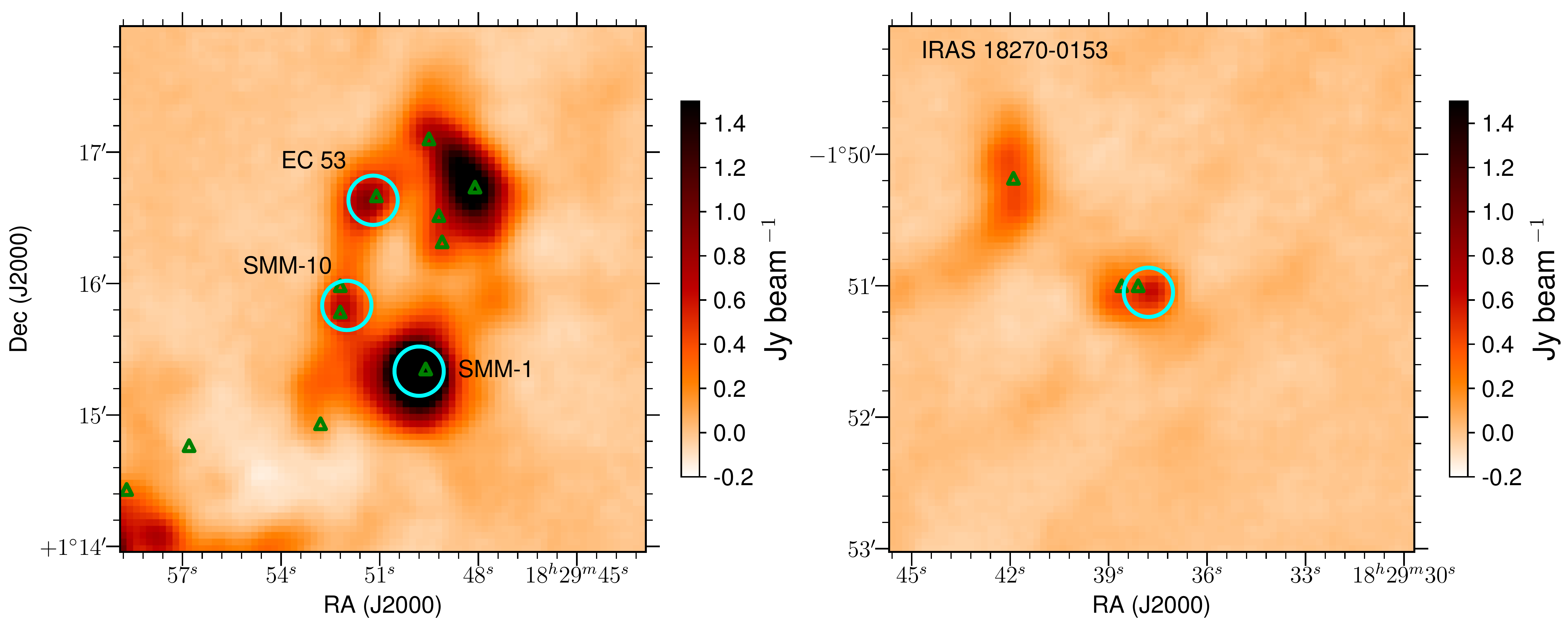}
}
\caption{Potential secular variable sources in the Serpens Main (left panel) and Serpens South (right panel) molecular clouds {overlaid on an 850\,$\mu m$ SCUBA-2 map}. The cyan circles indicate the {location of} the sources of interest and the green triangles indicate the positions of known protostars. Individual sources are discussed in Sections \ref{sec:EC53} (EC\,53), \ref{sec:SMM10} (SMM\,10), \ref{sec:IRAS18270} (IRAS\, 18270-0153), 
and \ref{sec:SMM1} (SMM\,1).}
\label{Fig:SerpThumb}
\end{figure}

\begin{figure}[htb]
\center{
\includegraphics[scale=0.25]{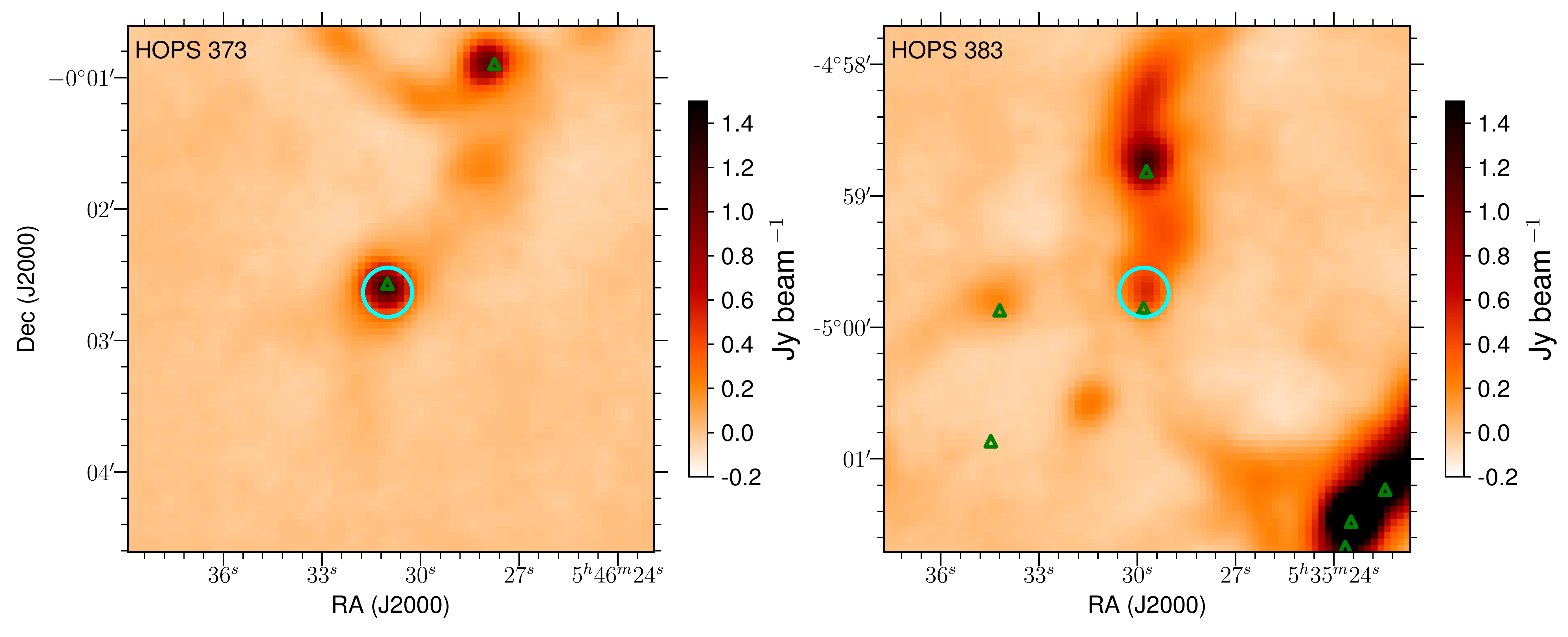}
}
\caption{Same as Figure \ref{Fig:SerpThumb} but for identified variable sources in the Orion B (left panel) and A (right panel) molecular clouds. Individual sources are discussed in Sections \ref{sec:HOPS373} (HOPS\,373) and \ref{sec:HOPS383} (HOPS383).}
\label{Fig:OrionThumb}
\end{figure}

\begin{figure}[htb]
\center{
\includegraphics[scale=0.22]{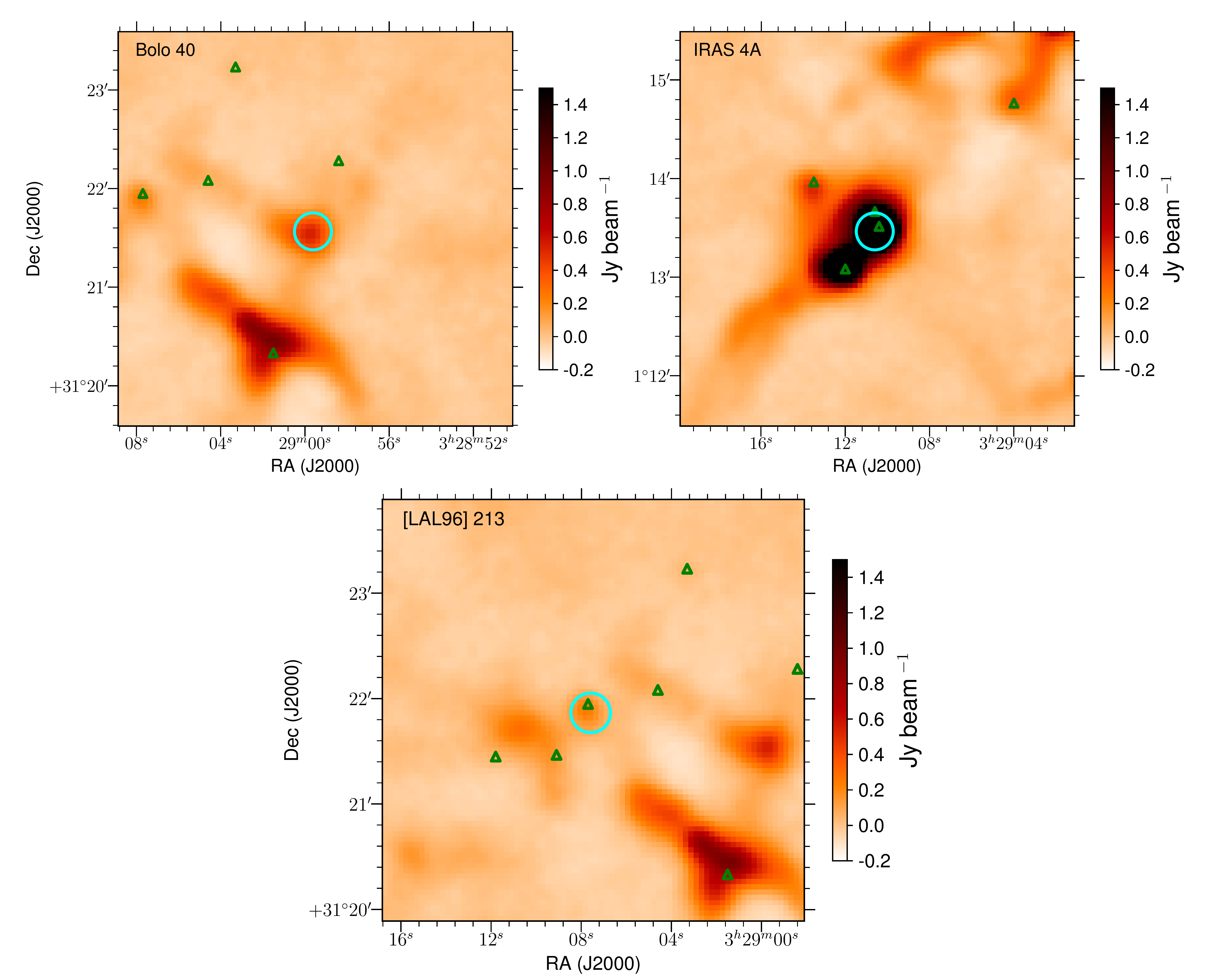}
}
\caption{Same as Figure \ref{Fig:SerpThumb} but for identified variable sources in the Perseus molecular cloud. Individual sources are discussed in Sections \ref{sec:BOLO40} (Bolo\,40), \ref{sec:IRAS4A} (IRAS\,4A), and \ref{sec:LAL96213} ([LAL96] 23).}
\label{Fig:PerseusThumb}
\end{figure}

\begin{figure}[htb]
\center{
\includegraphics[scale=0.22]{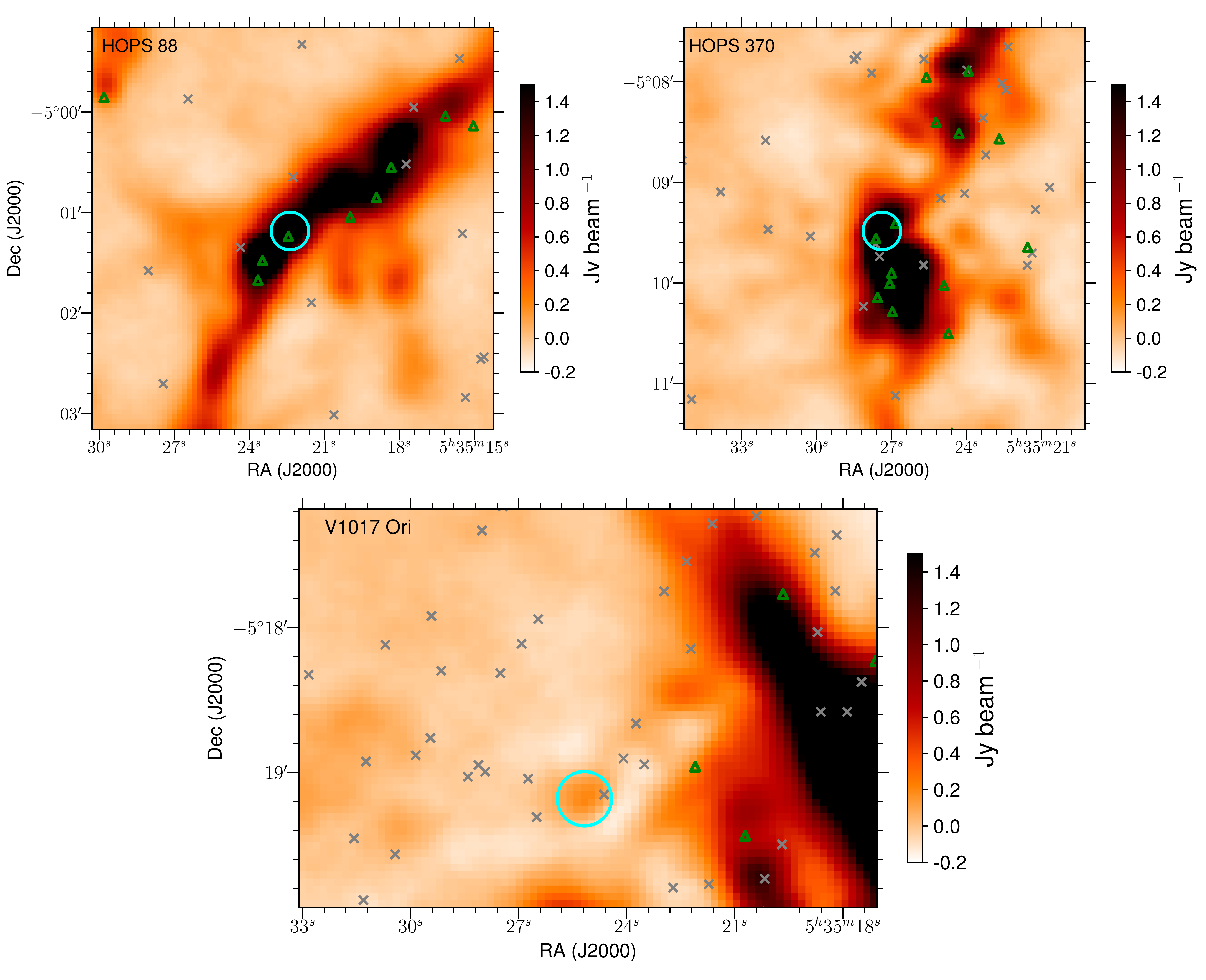}
}
\caption{Potential stochastic variable sources in the Orion A molecular cloud {overlaid on an 850\,$\mu m$ SCUBA-2 map}. The cyan circles indicate the location of the sources of interest, the green triangles indicate the positions of known protostars, and the grey exes indicate the positions of known disks. The lower bound of the V1017 Ori panel (bottom) is the edge of the Transient Survey OMC 2/3 region. Individuals sources are discussed in Sections \ref{sec:HOPS370} (HOPS\,370), \ref{sec:HOPS88} (HOPS\,88), and \ref{sec:V1017ORI} (V1017 Ori).
}
\label{Fig:OrionThumbStoc}
\end{figure}

\begin{figure}[htb]
\center{
\includegraphics[scale=0.15]{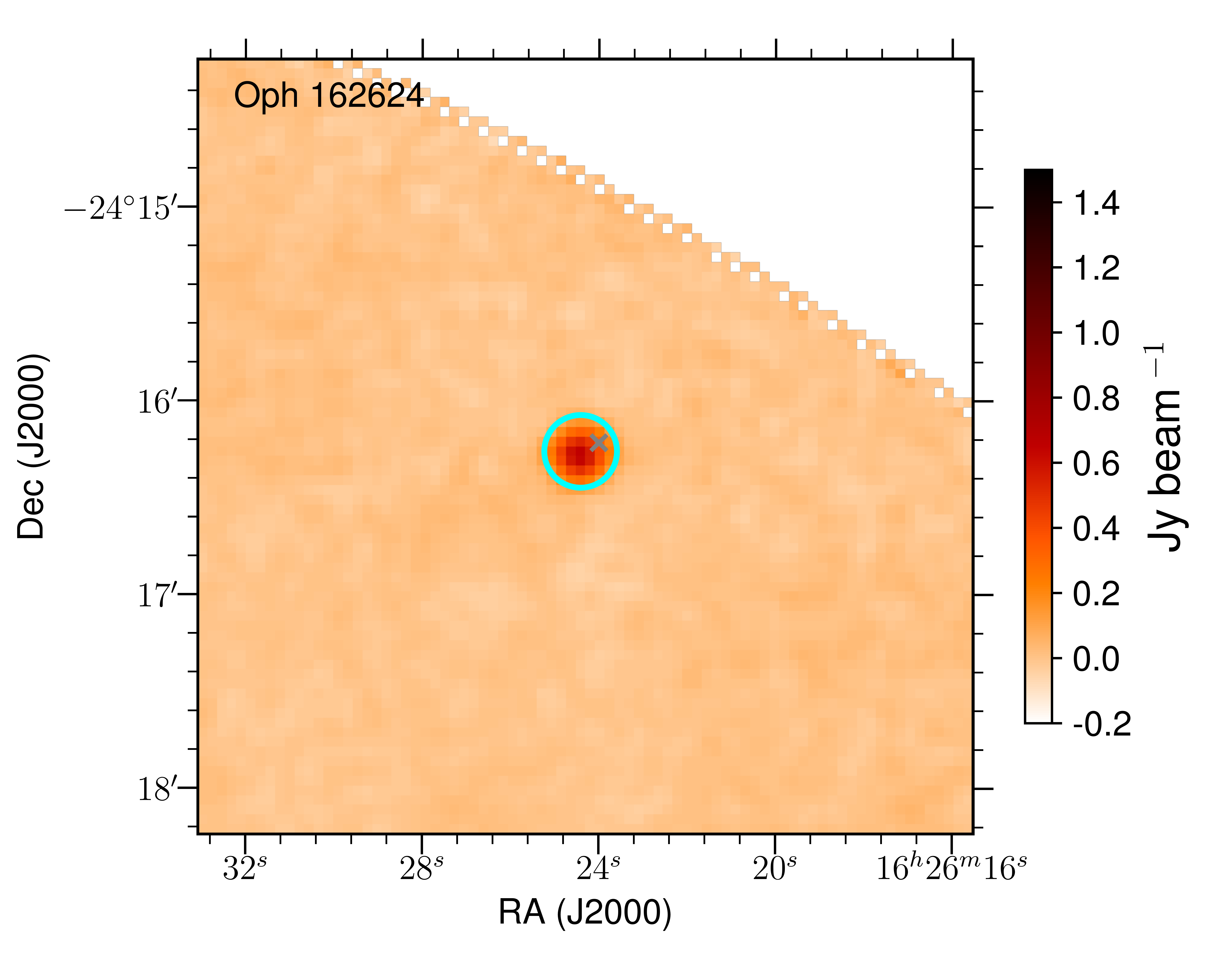}
}
\caption{Same as Figure \ref{Fig:OrionThumbStoc} but for the stochastic variable source in the Ophiuchus molecular cloud. Oph 162624 is discussed in Section \ref{sec:OPH162624}.}
\label{Fig:OphThumbStoc}
\end{figure}

Figures \ref{Fig:SerpThumb} through \ref{Fig:OphThumbStoc} provide finding charts for each candidate variable source to show the location relative to other structure in the field.  The following discussion also includes for each source a figure that shows the light curve and that describes the significance of secular and stochastic variability.   A full population analysis is beyond the scope of this paper but is anticipated after the completion of our 3-year survey.  In general, most variables show some signs of an outflow, and several have been previously identified as candidates of ongoing FUor-like outbursts.

\subsection{Secular and Stochastic Variable: EC\,53 (Serpens Main)}
\label{sec:EC53}

\begin{figure}[h]
\center{
\includegraphics[scale=0.7]{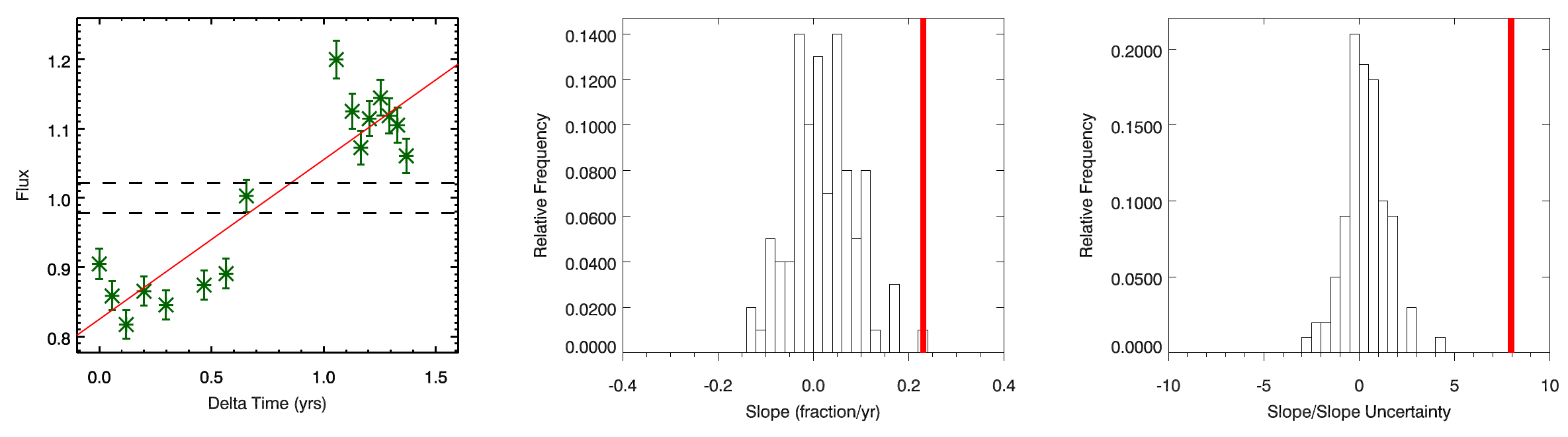}
}
\caption{EC\, 53 in Serpens Main. Left panel shows the sub-mm light curve over the observed epochs. Middle and right panels show histograms of the slope and slope uncertainty for 100 randomizations of the time-ordering of the flux measurements as well as vertical lines denoting the values derived for the observed light curve.}
\label{Fig:S1}
\end{figure}

EC\,53 in Serpens Main (see Figure\ \ref{Fig:SerpThumb}) was the first sub-mm variable revealed by the JCMT Transient Survey (see \citealt{yoo17} for a more complete description) and is a known near-IR periodic variable \citep{hodapp12}. This source has the largest stochastic variability measure in the Transient Survey (see Figure\ \ref{Fig:SD2} and Table\ \ref{Tab:SD}), as well as the strongest secular variability (see Table\ \ref{Tab:Secular}). 
%The sub-mm light curve for EC\,53 is presented in the left panel of Figure\ \ref{Fig:S1}, along with the best fit linear slope. 
Although the light curve should be periodic \citep[see][]{yoo17}, the partial phasing over which the source has been observed for this investigation makes it stand out as clear a linear variable (Figure\ \ref{Fig:S1}). 
%As additional checks on the quality of the linear fit, the center and right hand histograms show the measured slope and slope versus slope uncertainty, respectively, against 100 time-randomizations of the same flux measurements. In both histograms the time-ordered EC\,53 measurements lie well outside the randomized values. 
EC\,53 was not recovered as a variable source by \citet{mairs17b}, also due to the phasing of the GBS observations, although the GBS brightness is well described by the periodicity.  That the periodicity of EC 53 is identified here as secular variability underscores the degenerate interpretations of other secular variables as either long-term trends or periodic variables.
%They are, however, able to show that the GBS and Transient Survey observations if EC\,53 fit the periodic model.

% Figure 11

% Figure 12
\begin{figure}[htb]
\center{
\includegraphics[scale=0.7]{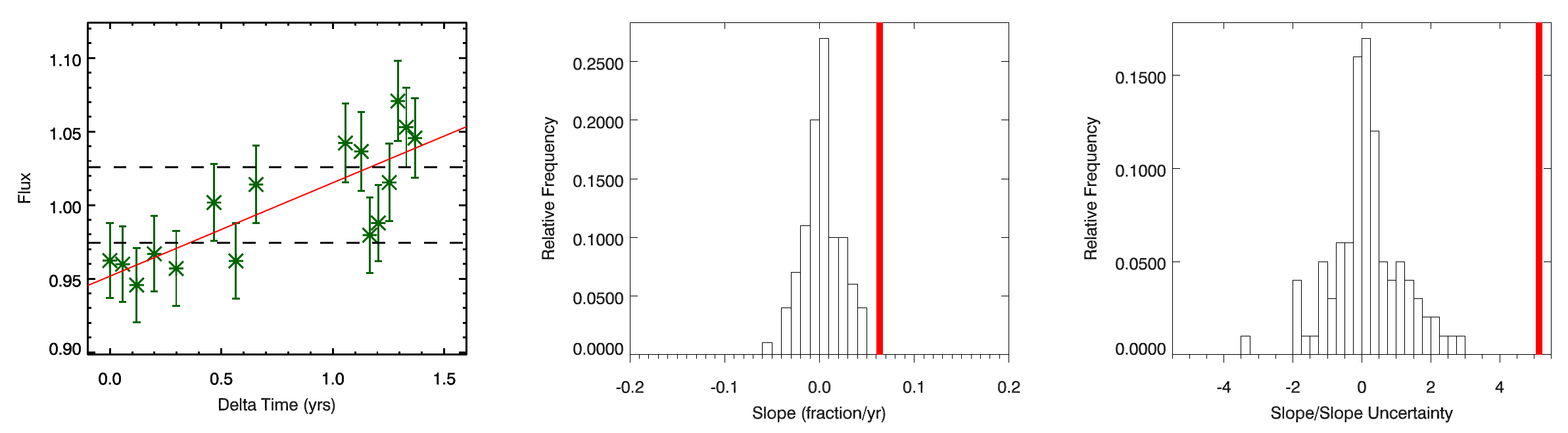}
\includegraphics[scale=0.7]{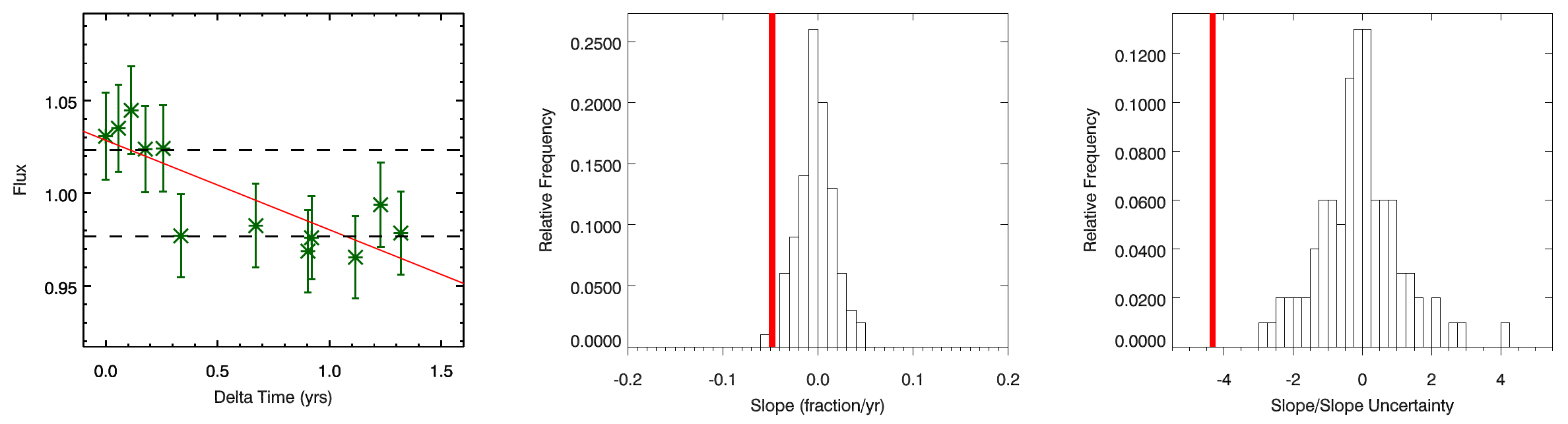}
\includegraphics[scale=0.7]{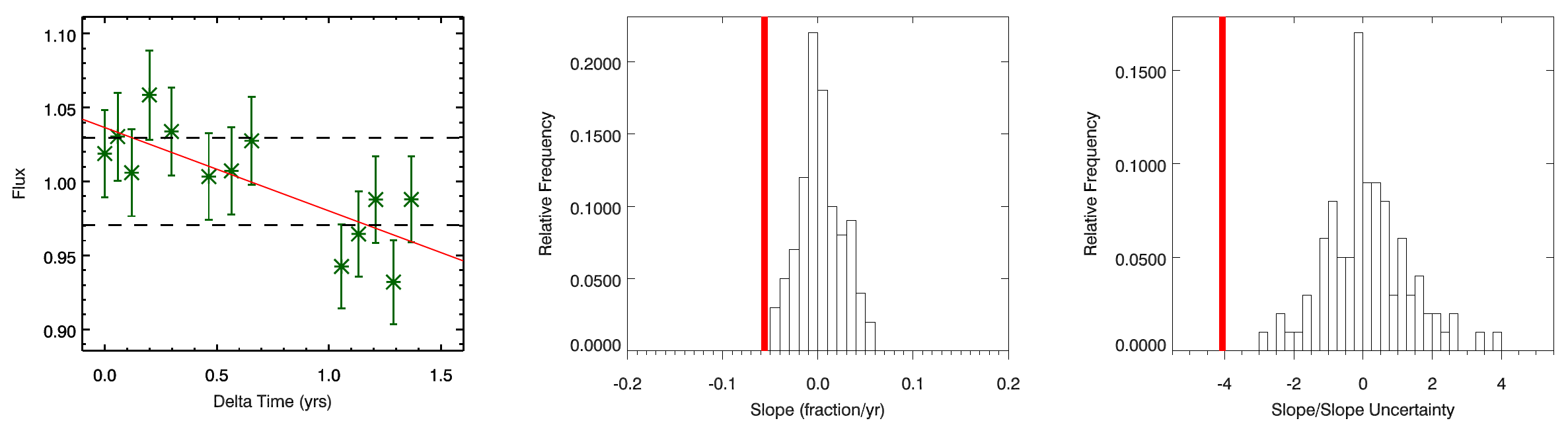}
\includegraphics[scale=0.7]{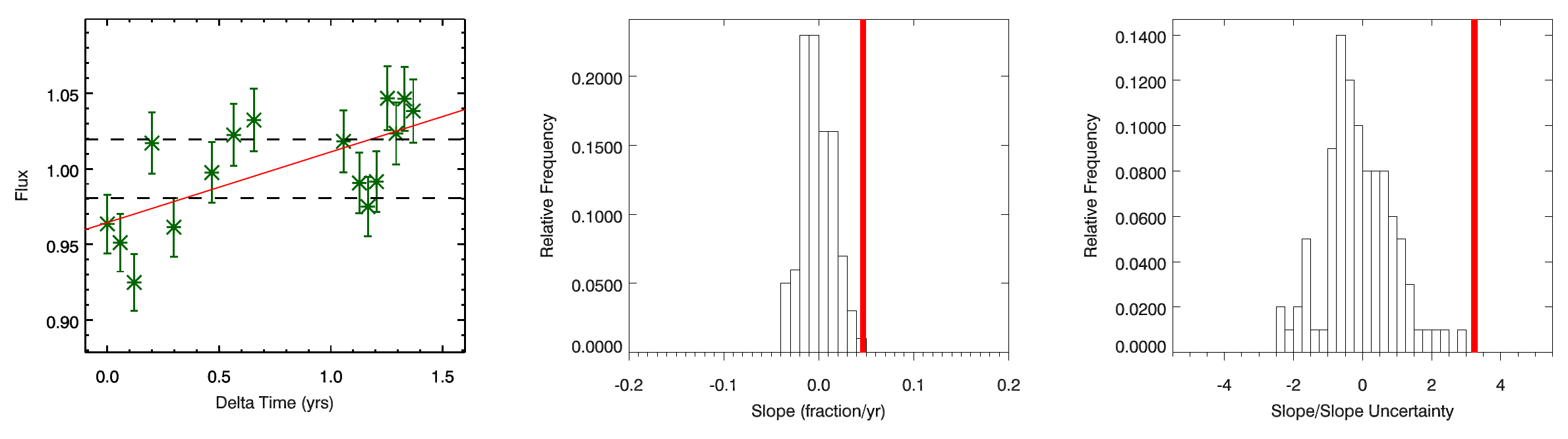}
}
\caption{Secular sub-mm variable sources. Top to bottom: SMM\,10 IR, in Serpens Main;  HOPS 373 in NGC\,2068; IRAS 18270-0153 in Serpens South; SMM\,1 in Serpens Main. Left panel shows the sub-mm light curve. Middle and right panels show histograms of the slope and slope versus slope uncertainty for 100 randomizations of the time-ordering of the flux measurements as well as vertical lines denoting the values derived for the observed light curve.}
\label{Fig:S2-5}
\end{figure}

\newpage
\subsection{Secular Variable: SMM\,10 (Serpens Main)}
\label{sec:SMM10}

SMM\,10 in Serpens Main (Figure\ \ref{Fig:SerpThumb})
%and the sub-mm light curve along with histograms showing the likelihood of the source having a significant secular variation are provided in Figure\ \ref{Fig:S2} . 
%This source 
has increased in brightness by $\sim 10$\% over the course of the survey (see Table \ref{Tab:Secular} and Figure\ \ref{Fig:S2-5}). Although the individual images of Serpens Main do not show any evidence of the variability of EC\,53 affecting other source measurements, monitoring this source over two full periods of EC\,53, the 3\,yr duration of the Transient Survey, will be helpful in disentangling any possible biasing by the presence of the nearby periodic variable. SMM\,10 was too faint to be included in the variability analysis by \citet{mairs17b}.

SMM\,10 is considered a Class 0/I object based on the red spectral energy distribution and bright sub-mm continuum emission \citep[e.g.][]{enoch09,dunham15}.  The source is associated with dense gas, which confirms that the protostar is embedded in an envelope \citep{lee14}.  However, mid-IR continuum and sub-mm CO 3--2 emission surveys have found no evidence of outflows \citep[e.g.][]{dionatos10,velusamy14}.

%SMM\,10 IR shows some evidence for an outflow (Velusamy+2014). 
%[Serpens Main: 18:29:52.00 +01:15:50.0: Ser SMM10 (NAME Serpens SMM 10 IR); not much. Velusamy+2014 Spitzer Outflow. Dense gas (Lee, Fernandez-Lopez, et al. 2014). No real outflows (Dionatos+2010; near-IR from Hodapp+1999). Strange Source.]

\subsection{Secular Variable: HOPS 373 (NGC 2068)}
\label{sec:HOPS373}

HOPS 373 in NGC 2068 (see Figure\ \ref{Fig:OrionThumb}) is a Class 0/I protostar, as measured from the red spectral energy distribution and bright sub-mm emission \citep[e.g.][]{sadavoy10,furlan16,kirk16}. As noted in Section\ \ref{Sec:DR}, this source is not included as a protostar in the original \citet{megeath12} catalogue but was noted by \citet{stutz13} as a PACS Bright Red Source (PBRS). The robust detection found here (Table\ \ref{Tab:Secular}) shows a step-like decline in brightness a few months after the start of our Transient Survey. HOPS 373 was also found to decline in brightness between the GBS and Transient Surveys but the significance of the decline, $\delta \sim 4$ (see \S \ref{Sec:Discussion} and Table\ \ref{Tab:Sources}), was just under the threshold for inclusion in the robust sample (see \citealt{mairs17b} for details). The sub-mm light curve and histograms showing the likelihood of this source having secular variability are shown in Figure\ \ref{Fig:S2-5}. 

HOPS 373 has a bright, wide-angle outflow with bright emission in highly-excited far-IR molecular lines \citep{tobin16,manoj16}, and may also be associated with a nearby HH-object with an H$_2$O maser \citep{haschick83,yu00}.

%[HOPS 373 (NGC 2068):  bright outflows (Manoj+2016 for PACS, Tobin+2016 for ground-based); also nearby maser.  Not much else.  K=16, near-IR may be outflow?;  deep near-IR imaging also available (Spezzi+2015)]

\subsection{Secular Variable: IRAS 18270-0153 (Serpens South)}
\label{sec:IRAS18270}

The source at 18:29:37.8 -01:51:03 in Serpens South is located $\sim 5^{\prime\prime}$ from IRAS 18270-0153, within our uncertainty beam (see Figure\ \ref{Fig:SerpThumb}).  This source has a robust decrease in brightness between the GBS and Transient Surveys \citep{mairs17b}. Here, we also find the source fading with time (Figure \ref{Fig:S2-5} and Table \ref{Tab:Secular}), and note that this fade may be either gradual or step-like. 

IRAS 18270-01530 is an embedded object with bright mm-continuum emission \citep[e.g.][]{maury11} and an outflow seen in near-IR H$_2$ and low-$J$ CO emission, but not in excited far-IR molecular emission \citep{zhang15,mottram17}.  IRAS 18270-01530 is considered an FUor candidate based on deep CO absorption in the K band \citep{connelley10}. Thus, the sub-mm fading may be consistent with a long-term decay of the outburst.

%[Serpens South:  18:29:37.8 -01:51:03.0:  Connelley+2008:  binary/multiplicity Connelley+2010:  18 29 38.92 −01 51 06.3 is labeled as an FUor (deep CO absorption in K-band); a little far from the source?;  Zhang+2015:  H2 outflow]

\subsection{Secular Variable: SMM 1 (Serpens Main)}
\label{sec:SMM1}

Serpens SMM\,1 in Serpens Main (see Figure\ \ref{Fig:SerpThumb}) is the brightest sub-mm source in Serpens.  Although SMM 1 has the highest likelihood of being a false positive among the five protostellar variables identified here (Table \ref{Tab:Secular}), the comparison between the GBS data and our Transient Survey yields a similar increase in brightness with time \citep{mairs17b}.
%has the largest likelihood of being a false positive of the five robust protostellar secular variables uncovered in this investigation . It is, however, the brightest source in Serpens and thus despite the possiblity of being a false-postive we consider it a good candidate for variability. Furthermore, this source is also found to be increasing in brightness with time between the  \citep{mairs17b}. 
The sub-mm light curve shown in Figure \ref{Fig:S2-5} suggests additional structure beyond a simple linear increase.

SMM1 is the prototypical intermediate-mass Class 0 star with a massive, $8$ M$_\odot$ envelope \citep[e.g.][]{enoch09smm1} and bolometric luminosity of $\sim 69$ L$_\odot$ \citep{dunham15}.  
The outflow from the source includes a jet, surrounded by an ionized outflow and a molecular outflow \citep[e.g.][]{goicoechea12,hull16}.  

%[Serpens Main 18:29:49.8:  Serpens SMM 1; prototypical intermediate-mass Class 0 protostar; very well studied outflow+jet (e.g., Hull+2016, 2017; Caratti o Garatti+2006 for near-IR H2).  No obvious evidence for outbursts in the jet.]

\subsection{Stochastic Variable: HOPS 370 (OMC $2/3$)}
\label{sec:HOPS370}

\begin{figure}[htb]
\center{
\includegraphics[scale=0.7]{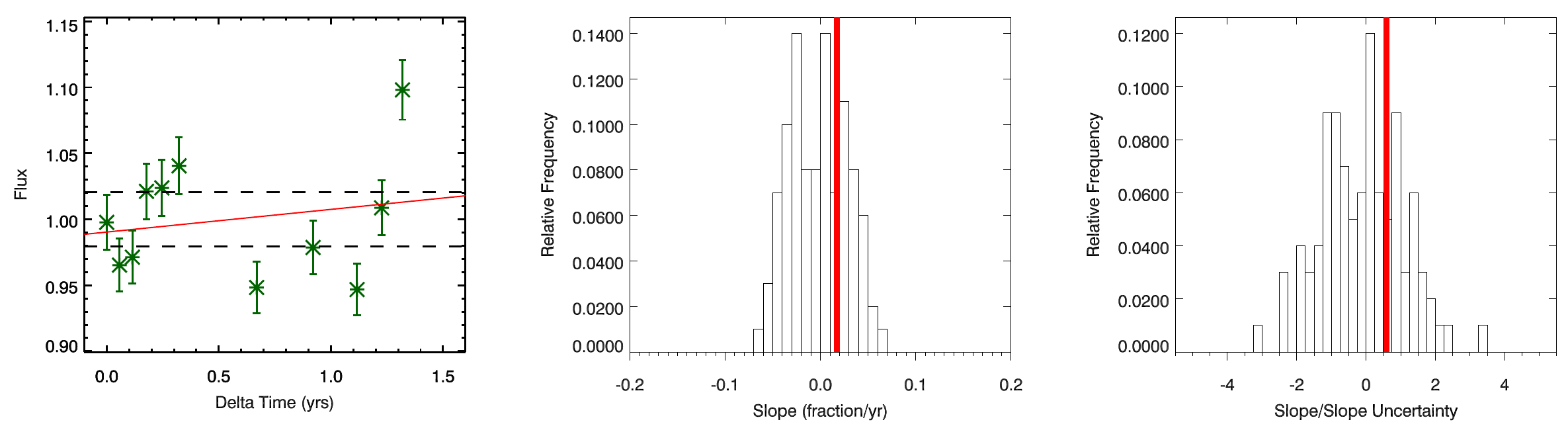}
\includegraphics[scale=0.7]{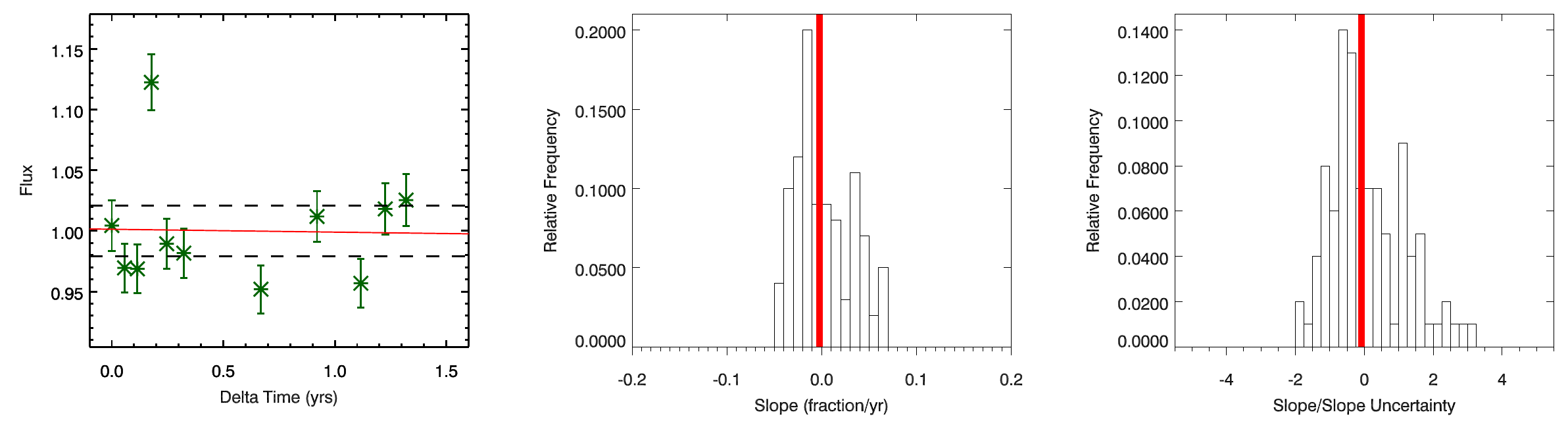}
\includegraphics[scale=0.7]{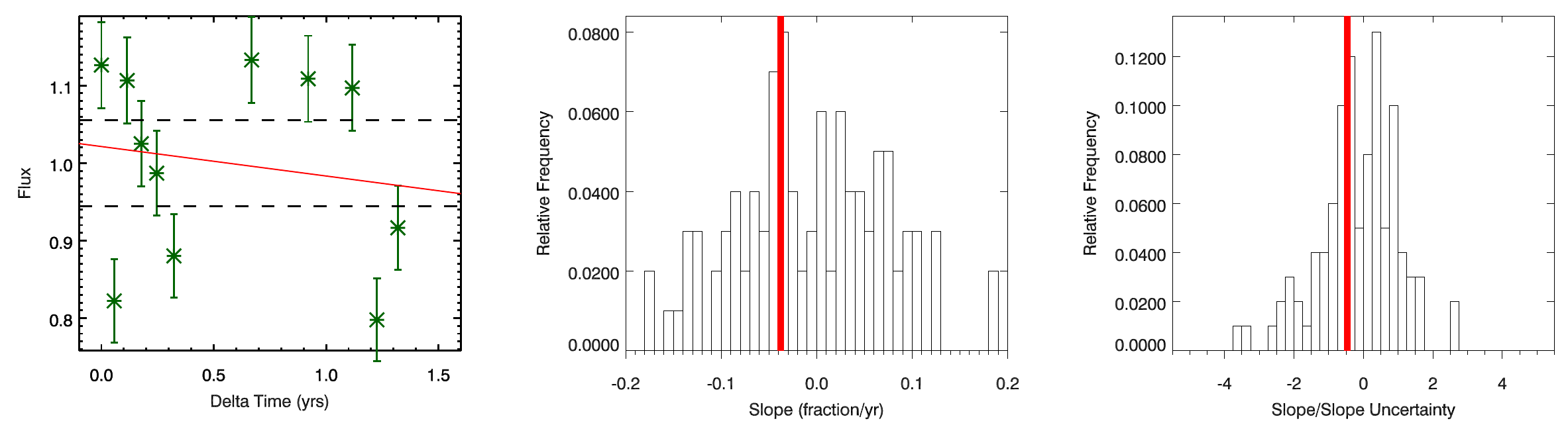}
\includegraphics[scale=0.7]{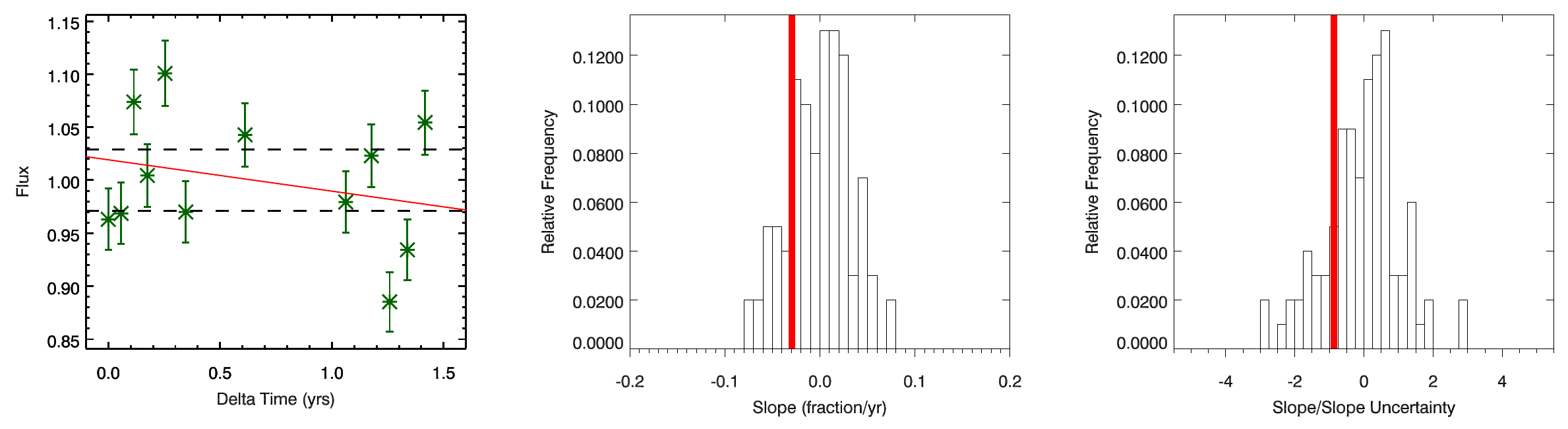}
}
\caption{Stochastic sub-mm variable sources. Top to bottom: HOPS 370 in OMC $2/3$; HOPS 88 in OMC $2/3$; V1017 Ori in OMC $2/3$; Oph 162624 in Ophiuchus.
Left panel shows the sub-mm light curve. Middle and right panels show histograms of the slope and slope versus slope uncertainty for 100 randomizations of the time-ordering of the flux measurements as well as vertical lines denoting the values derived for the observed light curve.}
\label{Fig:St2-5}
\end{figure}

HOPS 370 (OMC-2 FIR 3) in OMC $2/3$ (see Figure\ \ref{Fig:OrionThumbStoc}) is one of three potential stochastic variable protostars found in this investigation (Table \ref{Tab:SD}). The object is considered an intermediate-mass Class I object, with a bolometric luminosity of 361 L$_\odot$ \citep[e.g.][]{fischer17}.  Its sub-mm light curve (Figure \ref{Fig:St2-5}) shows that the standard deviation of the mean brightness of the source is dominated by a single epoch, during which the source appears to increase in brightness by $\sim 10$\%.  The protostar is likely associated with the optically-bright companion, V2467 Ori, located $\sim 3^{\prime\prime}$ away.

Lightcurves of V2467 Ori at 3.6 and 4.5 $\mu$m from {\it Spitzer}/IRAC show that the combined emission from these two sources is generally stable \citep{morales11}.  The source also has stable far-IR emission to within 10\% in six epochs obtained over 1.5 months with {\it Herschel}/ACS \citep{billot12}.  The source is associated with bright outflow emission \citep[e.g.][]{yu00,takahashi08,kang13}

%[OMC 23:  05 35 27.39 -05 09 29.0:  close to HOPS 370, aka OMC-2 FIR 3; [O I] jet from Gonzales-Garcia+2016; Billot+2012 variability at 70 (not 160); no near-IR variability (Morales-Calderon), but Billot detects secular variability; Takahashi+ also for outflows]

% Figure 13

\subsection{Stochastic Variable: HOPS 88 (OMC $2/3$)}
\label{sec:HOPS88}

HOPS 88 (OMC-3 MMS 5) in OMC $2/3$ (see Figure\ \ref{Fig:OrionThumbStoc}) is another of three potential stochastic variable protostars found in this investigation (Table \ref{Tab:SD}). Like HOPS 370, the sub-mm light curve of HOPS 88 (Figure \ref{Fig:St2-5}) shows that the standard deviation of the mean brightness of the source is dominated by a single epoch, albeit a different one than for HOPS 370, during which the source appears to increase in brightness by $\sim 10$\%.  HOPS 88 shows some variability in 3.6 and 4.5 $\mu$m continuum emission \citep{morales11}, and varied by 19\% in six epochs of far-IR continuum emission \citet{billot12}.  HOPS 88 is associated with dense gas and CO outflows \citep[e.g.][]{takahashi08,takahashi09}

%[OMC 23:  05 35 22.4:  HOPS 88 ([THT2013] OMC3-SMM 6):  Satoko has several papers.  dense gas, CO outflows (Takahashi+2008, 2009); morales-calderon+2011:  a little IRAC variability (but not all that much)]

\subsection{Stochastic Variable: V1017 Ori (OMC $2/3$)}
\label{sec:V1017ORI}

V1017 Ori in OMC $2/3$ (see Figure\ \ref{Fig:OrionThumbStoc}) is one of two potential stochastic variable disk sources found in this investigation (Table \ref{Tab:SD}).  The centroid of the sub-mm emission source is $10^{\prime\prime}$ from the optical/near-IR source, within a single sub-mm beam but too distant to be confident in the association.  The sub-mm light curve of V1017 Ori (Figure \ref{Fig:St2-5}) reveals that no individual measurement is responsible for the enhanced uncertainty in its measured brightness.

The spectral type M3.5 for V1017 Ori implies a mass of $\sim 0.2-0.3$ M$_\odot$ \citep{hillenbrand97,dario16}.  Optical spectra and photometry of V1017 also show strong H$\alpha$ emission, indicating ongoing accretion \citep[e.g.][]{manara12}.  Variability in the near-IR and in {\it Spitzer}/IRAC 3.6 and 4.5 $\mu$m lightcurves is likely caused by changes in the inner disk structure \citep{carpenter01,morales11}.

%[OMC23: 10" from V1017 Ori, a little too far. Morales-Calderon: IRAC variability of 0.2 mag; binary (? I don't think so) w/ disk. H97: M3.5, so low mass. H-alpha emission; Da Rio; Carpenter+2001: near-IR variability?. But, uncomfortably far from the source.]

\subsection{Stochastic Variable: Oph 162624 (Ophiuchus)}
\label{sec:OPH162624}

Oph 162624 (YLW 32; Elia 2-24) in Ophiuchus (see Figure \ref{Fig:OphThumbStoc}) is one of two potential stochastic variable disk sources found in this investigation (Table \ref{Tab:SD}). The sub-mm light curve (Figure \ref{Fig:St2-5}) reveals no individual measurement is responsible for the enhanced uncertainty in its measured brightness.

The brightness of the disk has made it a prominent object for high-resolution studies \citep[e.g.][]{andrews07}.  The star has an effective temperature and luminosity that imply a central mass of $\sim 1.5$, and the emission lines from the star indicate continued strong accretion \citep{manara15,rigliaco16}.  

%[OPH 162624: two SIMBAD sources with lots of citations:  	[SSG2006] MMS003, bright optical source Elia 2-24 (YLW 32).  Rigliaco GAIA-ESO:  Teff=4504, L=1.18, Manara:  log T=3.66, L=0.53, log Mdot=-7.11; continuous disk (no hole), Andrews+2010; outflows?;  Kirk+2017:  class II.  A little too faint for ASAS-SN (it pops up a few times, but unsure if detections are secure)]

\subsection{Literature Variable: HOPS 383 (OMC 2/3)}
\label{sec:HOPS383}

\begin{figure}[htb]
\center{
\includegraphics[scale=0.7]{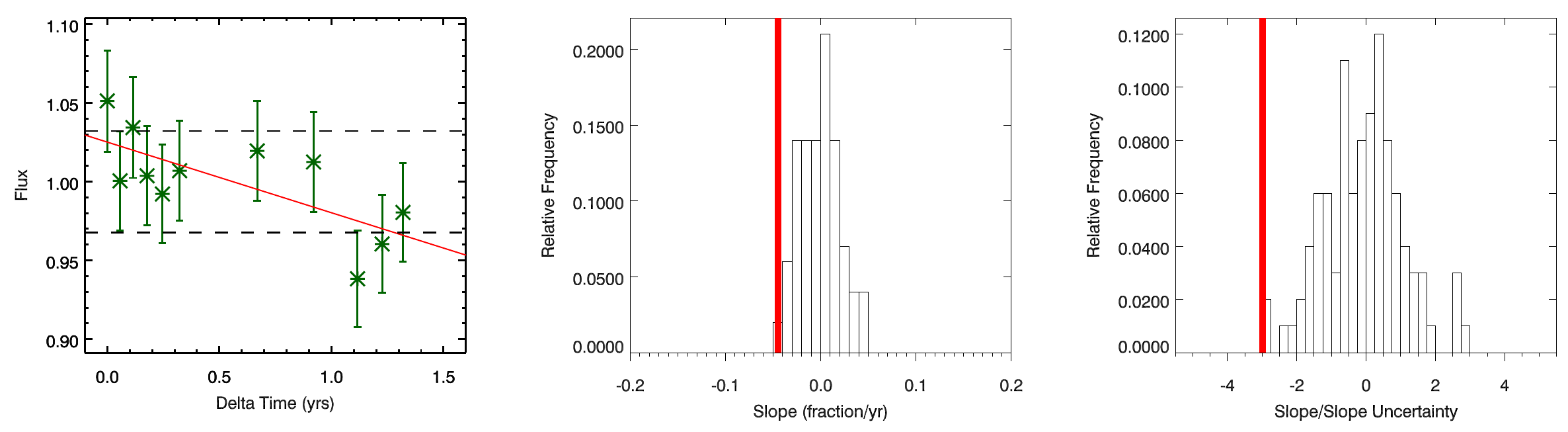}
\includegraphics[scale=0.7]{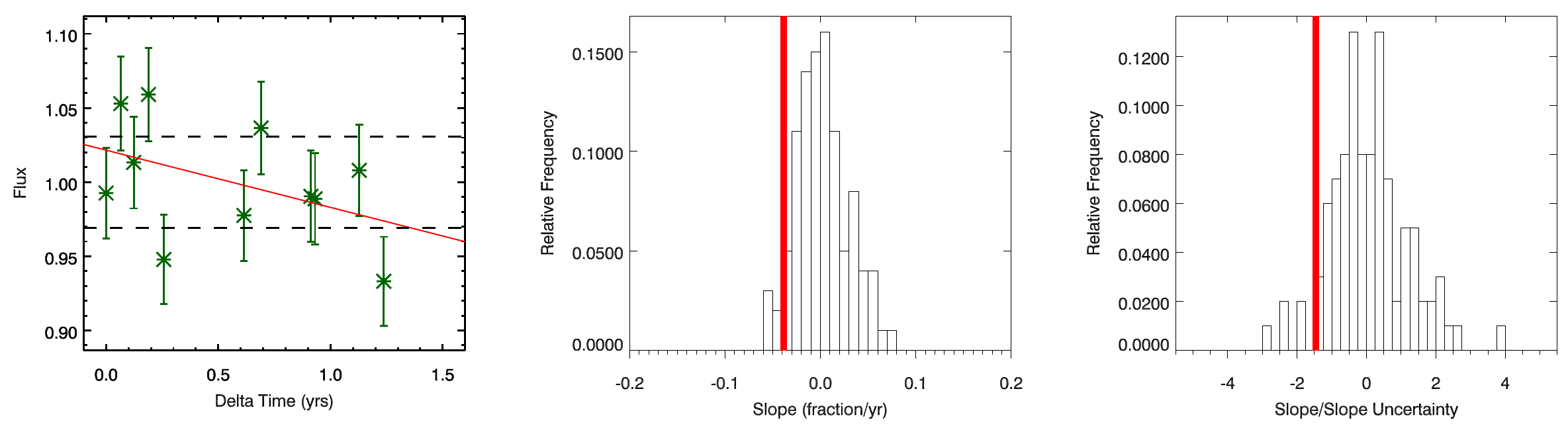}
\includegraphics[scale=0.7]{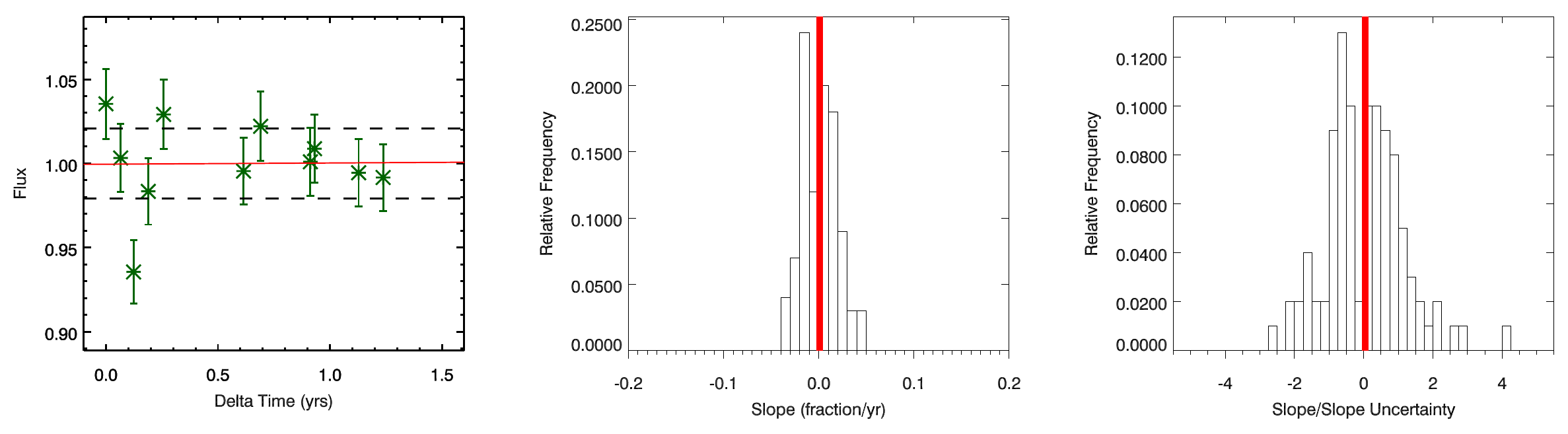}
\includegraphics[scale=0.7]{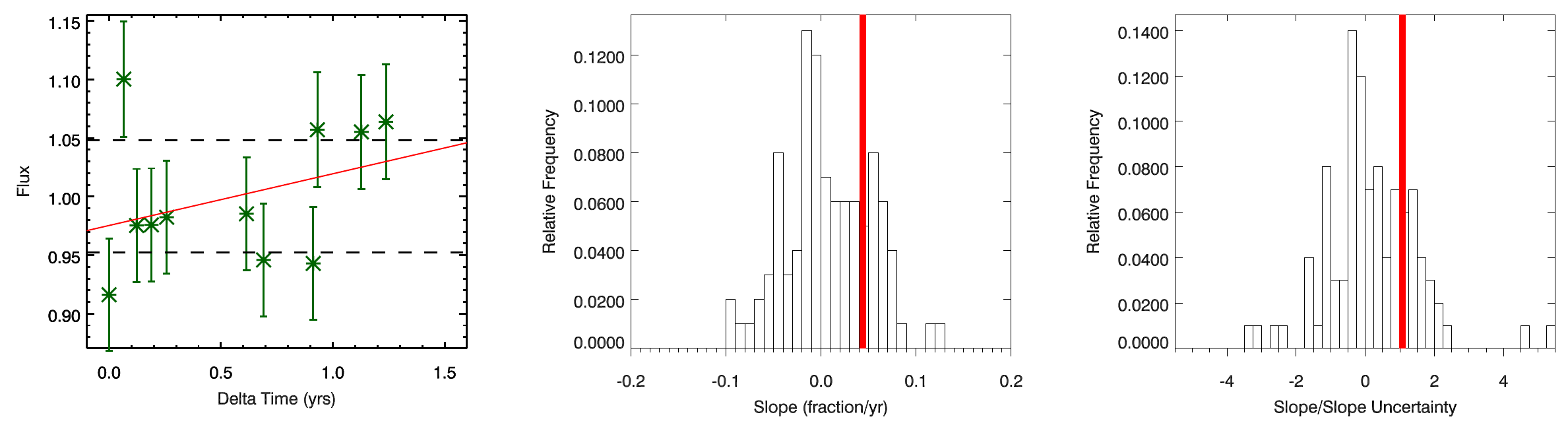}
}
\caption{Literature variable sources. Top to bottom: HOPS 383 in OMC 2/3; Bolo 40 in Perseus; NGC 1333 IRAS 4A in Perseus; LAL96 213 in Perseus.
Left panel shows the sub-mm light curve. Middle and right panels show histograms of the slope and slope versus slope uncertainty for 100 randomizations of the time-ordering of the flux measurements as well as vertical lines denoting the values derived for the observed light curve.}
\label{Fig:LS1-4}
\end{figure}

HOPS 383 in OMC 2/3 (at 5:35:29.67 -4:59:37.25; see Figure\ \ref{Fig:OrionThumb}) was found to decrease in brightness between the GBS Survey and the Transient Survey \citep{mairs17b}, albeit with a marginal significance. The sub-mm light curve (Figure \ref{Fig:LS1-4}) across only the Transient Survey also shows moderate evidence of a decrease, at a similar rate as found by \citet{mairs17b}. Previously, this object produced a remarkable outburst in IR and sub-mm wavelengths \citep{safron15}.  The fade in sub-mm continuum emission is in the same direction as the dramatic fade seen in the near-IR \citep{fischer17atel}, but is much more modest, indicating that either the near-IR suffers from increasing extinction or the time-delay between any change in the near-IR and sub-mm emission is much longer than expected.

% Figure 14

\subsection{Literature Variable: Bolo 40 (Perseus)}
\label{sec:BOLO40}

Bolo 40 in Perseus (3:28:59.86 31:21:33.09; see Figure\ \ref{Fig:PerseusThumb}) was found to decrease gradually, about 3 percent per year, between the GBS Survey and the Transient Survey \citep{mairs17b}. The sub-mm light curve (Figure \ref{Fig:LS1-4}) across only the Transient Survey shows only a hint of decline, which would be expected if source continued to dim in brightness slowly. Bolo 40 is the only potentially variable source discussed in this paper which is not known to be associated with either a protostar or disk.

\subsection{Literature Variable: NGC 1333 IRAS 4A (Perseus)}
\label{sec:IRAS4A}

NGC1333 IRAS 4A in Perseus (at 3:29:10.42 31:13:30.63; see Figure\ \ref{Fig:PerseusThumb}) was found to increase slowly, about 2 percent per year, between the GBS Survey and the Transient Survey \citep{mairs17b}. The sub-mm light curve (Figure \ref{Fig:LS1-4}) across only the Transient Survey, however, shows no clear evidence of a secular variation, as expected if the rise in brightness remained slow.

\subsection{Literature Variable: LAL96 213 (Perseus)}
\label{sec:LAL96213}

LAL96 213 in Perseus (at 3:29:07.66 31:21:54.05; see Figure\ \ref{Fig:PerseusThumb}) was found to decrease substantially between the GBS Survey and the Transient Survey \citep{mairs17b}. The sub-mm light curve (Figure \ref{Fig:LS1-4}) across only the Transient Survey, however, does not show any evidence of a secular decrease, suggesting that the rate of decline has tapered off over the last eighteen months. As shown in the right panels of Figure \ref{Fig:LS1-4}, the slope fit to the left panel is not significant.

\subsection{A Sample of Non-Variable Sources}
\label{sec:nonvariables}

\begin{figure}[htb]
\center{
\includegraphics[scale=0.7]{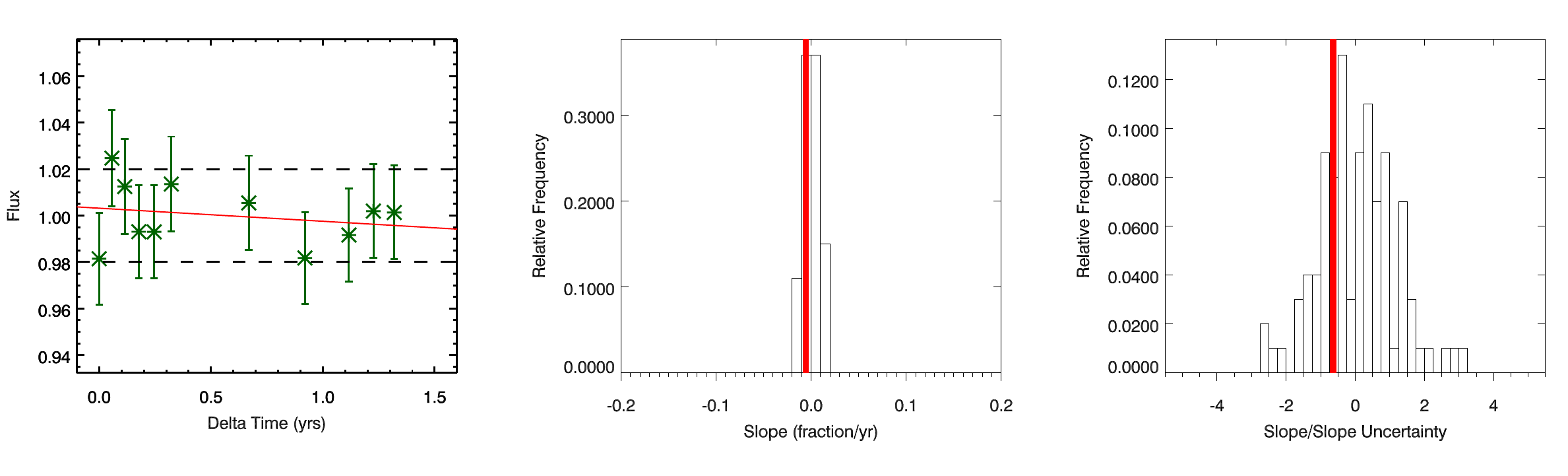}
\includegraphics[scale=0.7]{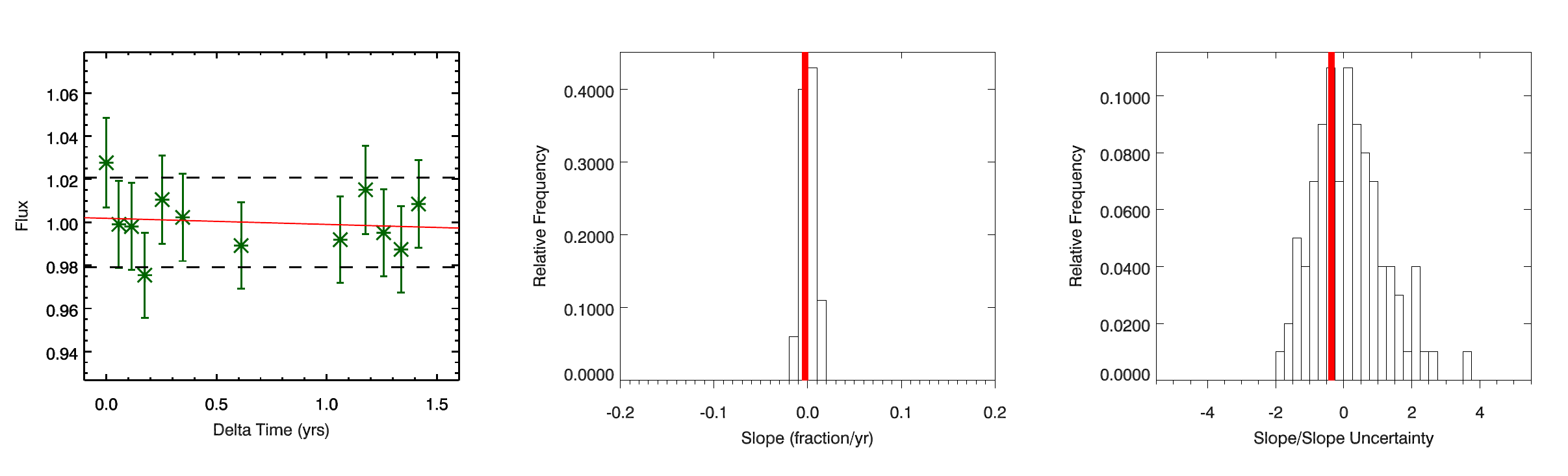}
\includegraphics[scale=0.7]{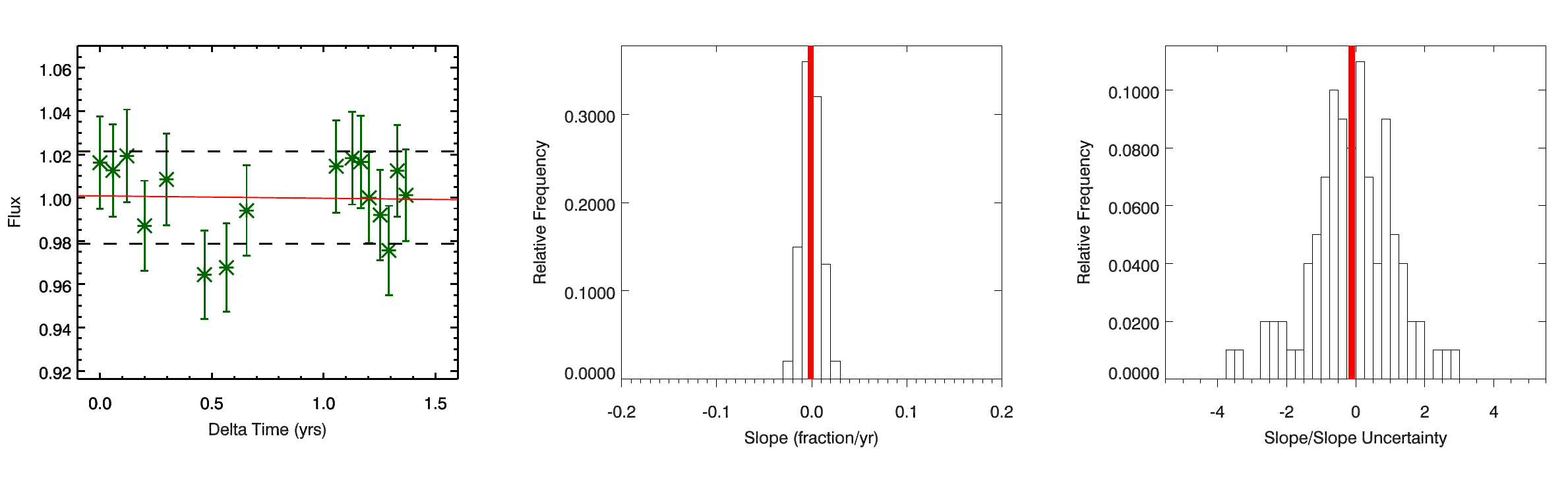}
}
\caption{Non-variable sources. Top to bottom: HOPS 64 in OMC\,2/3 (also known as OMC\,2 FIR\, 4); SMM J162628-24235 in the Ophiuchus Core; SH\,2-68\,N in Serpens Main.
Left panel shows the sub-mm light curve. Middle and right panels show histograms of the slope and slope versus slope uncertainty for 100 randomizations of the time-ordering of the flux measurements as well as vertical lines denoting the values derived for the observed light curve.}
\label{Fig:NV1}
\end{figure}

Figure \ref{Fig:NV1}  provides light curves for three sources that have not been observed to vary during the Transient Survey so far. These are not random objects, but rather the brightest sources in OMC\,2/3 [HOPS 64 also known as OMC\,2 FIR\,4; \citet{furlan14}] and the Ophiuchus Core [SMM J162628-24235; \citet{jorgensen08}], as well as the second brightest source in Serpens Main [SH\,2-68\,N; \citet{winston07}].  Note that the brightest source in Serpens Main, SMM\,1, was found to vary secularly (\S \ref{sec:SMM1}). These sources are presented as examples of how flat the light curves can be when the sources are bright and the measured uncertainties are small (see, for example, the middle panel in each figure which plots the range of slopes allowed by randomizing the time-order of the measurements).

% Figure 

\section{Discussion}
\label{Sec:Discussion}
Our search for secular changes in sub-mm brightness over the lifetime of the JCMT Transient Survey 
has uncovered five robust variables (see Table\ \ref{Tab:Secular}). Another source, HOPS 383, is likely to be a secular sub-mm variable (\S \ref{sec:HOPS383}). These numbers can be compared against the total number of protostars observed with sub-mm peaks $> 0.35\,$Jy\,bm$^{-1}$ (51; Table\ \ref{Tab:SourceList}). We, therefore, find that few protostars vary with large, $> 10\%$, amplitudes in the sub-mm continuum over year timescales, but $ \sim 10\%$, vary by $\sim 5\%$ over a year. This later fraction is in agreement with the result obtained by \citet{mairs17b}, who compared the mean Transient Survey brightnesses of sources against the mean brightness measured from data obtained two to four years earlier by the JCMT Gould Belt Survey \citep{ward-thompson07}.  As such, it would appear that moderate variability of protostars at sub-mm wavelengths is relatively common. These early results will be considerably strengthened over the lifetime of the Transient Survey, as discussed in \S \ref{Sec:Slope:Toy}. Significant numbers of sources with linear brightness changes as small as $2\%$ per year should be directly observable at that time.

All five secular variables identified in this paper are protostellar, despite our sample including 
twice as many bright starless cores as protostars (see \S\ \ref{Sec:Slope}). While the brightness distribution of starless cores is skewed toward lower values than the protostellar sample, the bright end contains a similar number of starless cores as protostars (see Figure \ref{Fig:slopes}).  Thus, to first order the two samples are quite similar over the range of brightnesses for which secular variability was uncovered in the protostellar sample. As such, assuming that both the protostar and prestellar samples have the same underlying variability properties, the likelihood of all five sources being drawn from the protostellar sample is about 3\,\%. While not impossible, it is much more likely that the prestellar sample is not varying at the same level as the protostellar sample.  This difference is expected because prestellar cores do not have any known source of energy to provide a time-dependent brightness at sub-mm wavelengths on month to year timescales.

The lack of strong stochastic variables within the present sample, excluding EC\,53, suggests that significant brightness changes $> 10\%$ rarely occur over very short (monthly--to-yearly) timescales for the typical protostar or disk source. Indeed, the only source observed to have a large standard deviation in its brightness measurements, EC\,53, is best fit as an eighteen month periodic variable with a large amplitude, rather than as a randomly varying source. Perhaps more interesting are the possible additional periodic variables within the present data set with low amplitudes below our present detection limit, which will require time-domain Fourier analyses to uncover. For such sources, a greater number of epochs and a larger time range to compare against will be extremely beneficial.

%chopped from body of text
\begin{deluxetable}{llccccr}
\tablecolumns{7}
\tablewidth{0pc}
\tablecaption{Comparison of Identified Variable Sources \label{Tab:Sources}}
\tablehead{
\colhead{} &
\colhead{}&
\multicolumn{2}{c}{Transient Analysis}&\multicolumn{2}{c}{Transient-GBS Analysis}&
\colhead{}\cr
\colhead{Region}&
\colhead{Name} & 
\colhead{$(S/\Delta S)$\tablenotemark{a}}  &
\colhead{$S$} & 
\colhead{$\delta$\tablenotemark{b}}  &
\colhead{$S$} & 
\colhead{Comment} \cr
\colhead{}&
\colhead{} & 
\colhead{}  &
\colhead{(yr$^{-1}$)} & 
\colhead{}  &
\colhead{(yr$^{-1}$)} & 
\colhead{} \cr
}
\startdata
Serpens M&EC\,53& 7.9& 0.28& NA& NA& See Section \ref{sec:EC53}.\\
\hline
Serpens S&IRAS 18270-0153& 4.1& -0.05& 11.81& -0.04& Strong detection by both analyses.\\
NGC 2068& HOPS 373& 4.3& -0.05& 5.34& -0.04& Strong detection by both analyses.\\
Serpens M&SMM-1&3.2&0.05&6.85& 0.02& Strong detection by both analyses.\\
OMC\,2/3& HOPS 383& 3.0&-0.04& 4.17& -0.03& Moderate detection by both analyses.\\
\hline
NGC\,1333& Bolo 40& 1.5& -0.04& 7.99& -0.03&Only source not identified with protostar.\\
NGC\,1333& IRAS 4A& - & - & 7.66& 0.02& Not detected by present analysis.\\
NGC\,1333&[LAL96] 213& -& -& 8.31& -0.09& Not detected by present analysis.\\
\hline
Serpens M& SMM 10& 5.1& 0.07& NA& NA&Source too faint for Transient-GBS detection.\cr
\enddata
\tablenotetext{a}{Sources with $|S/\Delta S| > 4$ are robust against false-positives within the entire ensemble (see \S \ref{Sec:Slope}). Those sources with $|S/\Delta S | \geq 3$\\ are strong candidates when treated as a special case.}
\tablenotetext{b}{Sources with $\delta > 5$ are robust against false-positives within the entire ensemble (see Mairs et al.\ 2017a). Those sources with\\ $\delta > 4$ are strong candidates when treated as a special case.}\end{deluxetable}

Many of the secular variables identified in this paper show light-curves that appear to be more complicated than a simple linear variation in time, see \S \ref{Sec:Individual}. Nevertheless, three of the five sources identified here as secular variables are also seen to strongly vary by \citet{mairs17b} as shown in Table\ \ref{Tab:Sources}.  One of the two non-detections, EC\, 53, was excluded from the Mairs et al.\ variable source list precisely because of the large standard deviation of its individual brightness measurements. The other source, SMM\,10, was too faint for the Mairs et al.\ analysis.  Similarly, of the five robust variable detections found by Mairs et al.\ ($\delta > 6$ in Table\ \ref{Tab:Sources}), we recover two as robust secular variables and one more as possibly varying. The larger time separation, two to four years, for the Mairs et al.\ analysis allowed for the recovery of sources varying with lower amplitudes than detectable over the first eighteen months of the Transient Survey and this likely accounts for the poorer match between sources identified by \citet{mairs17b} and this paper. An additional ninth source, the known sub-mm variable HOPS 383, is found to be a likely-dimming variable by both analyses. For all of these sources, the sign of the brightness slope (rising or dimming) is always the same between the two analyses, when measurable (see Table\ \ref{Tab:Sources}). The agreement between these two independent investigations suggests that the assumption of a long-term, multi-year, timescale underlying the light curve is reasonable. As with the stochastic discussion above, future observations of these sources will help identify any quasi-periodic underlying nature.

It is important to recognize that the sub-mm continuum emission is likely responding to the change in the dust temperature within the protostellar envelope and accretion disk, and is not necessarily directly proportional to the underlying change in accretion luminosity of the central source \citep{johnstone13,yoo17}. Detailed modeling is required to determine the exact relationship between the change in sub-mm brightness and the change in the accretion luminosity, taking into account the range of dust temperatures in the disk and envelope as well as the importance of external heating on the outer envelope. \citet{mairs17b} consider these issues and conclude that the amplitude of the change in the accretion rate should scale somewhere between the amplitude of the sub-mm brightness change and the fourth-power of the sub-mm brightness change. Thus, in rough agreement with the numbers computed by Mairs et al., we find that about 10\% of the protostars in the Transient Survey are undergoing mass accretion variations between 5\% and 20\% over the course of a year. 
This value lies intermediate between the numerical results for accretion instability driven by magneto-rotation \citep{bae14} and those driven by large-scale modes within a gravitationally-unstable disk \citep{vorobyov10}, both of which are presented by \citet{herczeg17}.
However, neither of these models was intended to be used for such short timescale measurements. Nevertheless, this discussion show that results from the Transient Study will help to constrain  the next generation of accretion disk numerical models.  

This interpretation assumes that the sub-mm emission variability is caused by a change in the protostellar luminosity.  However, the two stochastic variables with single discrepant points may instead have individual bright epochs caused by radio synchrotron flares \citep[e.g.][]{bower03,forbrich17}.  {In these situations the spectral index in the optically thick regime would be flat or inverted such that the flux at centimeter wavelengths would be comparable or lower than in the sub-mm. As an example, the YSO GMR A in Orion was observed to brighten to 100 mJy at 3\, mm wavelengths \citep{bower03}.} Since these magnetic flares typically last for hours, they would be unable to explain the other sub-mm variables identified in this paper.

A curious result from the present survey is the number of variable sources found in Serpens Main. Three of the five secular variables are uncovered in this region. All are found to increase in brightness with time during the epochs observed (although  EC\,53 is actually periodic and is poorly fit by a linear rise in brightness, see Figure \ref{Fig:S1}). The individual epochs for the other two Serpens Main sources do not appear to be influenced by the large fluctuations seen in EC\,53. Indeed, EC\,53 is significantly fainter than SMM\,1 (see \S \ref{Tab:Secular}). The Serpens Main region has five independent calibrator sources, all of which have flat light curves \citep{mairs17}, including the source SH\,2-26\,N whose light curve is shown in Figure\ \ref{Fig:NV1}. 

Finally, an obvious extension of the JCMT Transient Survey is to use the ALMA sub-mm array to search for variations in the sub-mm morphology and brightness of deeply embedded protostars \citep[see, e.g.][]{hunter17}. ALMA's high spatial resolution images will yield important information on changes to the physical and chemical conditions within both the inner disk region and the jet/outflow emanating from these sources. Furthermore, the ALMA archive already contains many examples where protostars have been observed multiple times in the same wavelength band, with at least one continuum window and with a similar spatial resolution. The required continuum observations for these bright targets are extremely short, even to reach signal to noise ratios of greater than 300, and thus most of the archived measurements sets have the necessary sensitivity. The ability to achieve a precise relative calibration across multiple epochs of individual targets, however, is complicated both by variations in antenna configuration (and source rotation on the sky) and the stability of known calibrator sources.  These complications may be minimized with dedicated observations obtained with similar configurations and in the same bandpass.  A first epoch of a pilot survey has been obtained in Cycle 3 (PI D. Johnstone) and is now being used to investigate best practices for data reduction and analysis to optimize the precision of the flux calibration.

\section{Summary}
\label{Sec:Summary}
The JCMT Transient Survey \citep{herczeg17} is halfway through its three year monthly-monitoring of eight nearby star-forming regions at sub-mm wavelengths. In this paper we have analyzed the first eighteen months of Transient Survey data to uncover initial statistics on the rate of variability among protostars and disk sources. We find that only one protostar, EC\,53 in Serpens Main \citep{yoo17}, shows large $> 10\%$ variations in its sub-mm brightness over time and note that these variations are due to an eighteen month periodicity. Two additional protostars show potential stochastic variations, each having a single epoch with a strong statistical outlier measurement. Additional observations are needed to determine the significance of these rare events. We further find that five of the 150 brightest sub-mm peaks are well fit by a linear variation in brightness. All five of these sources are associated with known protostars, representing $10\%$ of the protostellar sample within the 150 bright sub-mm peaks. Three of these secular variables are also found to be varying by \citet{mairs17b} through their independent analysis of sub-mm brightness changes across two to four years. None of the 100 bright peaks unassociated with protostars are found to vary. Finally, analysis of a toy model for the underlying distribution of secular brightness variations within the protostellar sample reveals that a Gaussian distribution of fractional brightness change per year of $\sigma_{S_t} = 0.005$ would be unobservable at present, while $\sigma_{S_t} = 0.02$ is ruled out.

\acknowledgments{
The authors thank both the anonymous referee and Neal Evans for suggestions that significantly enhanced the readability of this paper. The authors gratefully acknowledge the hospitality of the Kavli Institute at Peking University where a majority of the analysis and paper writing was conducted while DJ was a Visiting Scholar. DJ is supported by the National Research Council of Canada and by an NSERC Discovery Grant.  SM was partially supported by the Natural Sciences and Engineering Research Council (NSERC) of Canada graduate scholarship program. GH is supported by general grant 11473005 awarded by the National Science Foundation of China. MK is supported by Basic Science Research Program through the National Research Foundation of Korea(NRF) funded by the Ministry of Science, ICT \& Future Planning (No.\ NRF-2015R1C1A1A01052160).

The authors wish to recognize and acknowledge the very significant cultural role and reverence that the summit of Maunakea has always had within the indigenous Hawaiian community. We are most fortunate to have the opportunity to conduct observations from this mountain. 

The JCMT is operated by the East Asian Observatory on behalf of The National Astronomical Observatory of Japan, Academia Sinica Institute of Astronomy and Astrophysics, the Korea Astronomy and Space Science Institute, the National Astronomical Observatories of China and the Chinese Academy of Sciences (Grant No. XDB09000000), with additional funding support from the Science and Technology Facilities Council of the United Kingdom and participating universities in the United Kingdom and Canada. The identification number for the JCMT Transient Survey under which the SCUBA-2 data used in this paper can be found is M16AL001.

The authors thank the JCMT staff for their support of the GBS team in data collection and reduction efforts. The Starlink software \citep{currie2014}  is supported by the East Asian Observatory. 

This research has made use of NASA's Astrophysics Data System and the facilities of the Canadian Astronomy Data Centre operated by the National   Research Council of Canada with the support of the Canadian Space Agency. This research used the services of  the Canadian Advanced Network for Astronomy Research (CANFAR) which in turn is supported by CANARIE,  Compute Canada, University of Victoria, the National Research Council of Canada, and the Canadian Space Agency.}

\facility{JCMT (SCUBA-2) \citep{holland13}}
\software{Starlink \citep{currie2014}, SMURF \citep{jenness2013}, CUPID \citep{berry2013}, Astropy \citep{astropy},  APLpy \citep{aplpy}, Matplotlib \citep{matplotlib}}

\bibliographystyle{apj}
\bibliography{ms}

\end{document}